\documentclass[aps,prd,amsmath,two column,amssymb,showpacs]{revtex4-1}
\usepackage{amssymb}
\usepackage{txfonts}
\usepackage{mathbbold}
\usepackage{amsfonts}
\usepackage{mathrsfs}
\usepackage{epsfig,bm,dcolumn}
\usepackage{graphicx}
\usepackage{color}
\usepackage{amsmath}

\begin{document}

\title{Three-flavor Nambu--Jona-Lasinio model at finite isospin chemical potential}

\author{Tao Xia,$^{1}$ Lianyi He,$^{2}$ and Pengfei Zhuang$^{1}$}
\address{1 Department of Physics, Tsinghua University, Beijing 100084, China\\
2 Frankfurt Institute for Advanced Studies, J. W. Goethe University, 60438 Frankfurt am Main, Germany}

\date{\today}

\begin{abstract}
QCD at finite isospin chemical potential $\mu_{\text I}$ possesses a positively definite fermion determinant
and the lattice simulation can be successfully performed. While the two-flavor effective models may be
sufficient to describe the phenomenon of pion condensation, it is interesting to study the roles of the
strangeness degree of freedom and the U$_{\rm A}(1)$ anomaly. In this paper, we present a systematic study
of the three-flavor Nambu--Jona-Lasinio model with a Kobayashi-Maskawa-'t Hooft (KMT) term that mimics the
U$_{\rm A}(1)$ anomaly at finite isospin chemical potential. In the mean-field approximation, the model
predicts a phase transition from the vacuum to the pion superfluid phase, which takes place at $\mu_{\rm I}$
equal to the pion mass $m_\pi$. Due to the U$_{\rm A}(1)$ anomaly, the strangeness degree of freedom couples
to the light quark degrees of freedom and the strange quark effective mass depends on the pion condensate.
However, the strange quark condensate and the strange quark effective mass change slightly in the
pion superfluid phase, which verifies the validity of the two-flavor models. The effective four-fermion
interaction of the Kobayashi-Maskawa-'t Hooft term in the presence of the pion condensation is constructed.
Due to the U$_{\rm A}(1)$ anomaly, the pion condensation generally induces scalar-pseudoscalar interaction.
The Bethe-Salpeter equation for the mesonic excitations is established and the meson mass spectra are obtained
at finite isospin chemical potential and temperature. Finally, the general expression for the topological
susceptibility $\chi$ at finite isospin chemical potential $\mu_{\rm I}$ is derived. In contrast to the
finite temperature effect which suppresses $\chi$, the isospin density effect leads to an enhancement of $\chi$.
\end{abstract}


\maketitle

\section{Introduction}\label{sec1}
Good knowledge of quantum chromodynamics (QCD) at finite temperature and density is
important for us to understand a wide range of physical phenomena. To understand the
evolution of the early Universe in the first few seconds, we need the nature of the QCD
phase transition at temperature $T\sim170$MeV and nearly vanishing baryon density. On
the other hand, to understand the physics of compact stars, we need the knowledge of the
equation of state and dynamics of QCD matter at high baryon density and low temperature.
Lattice simulation of QCD at finite temperature and vanishing density has been
successfully performed. However, at large baryon density the lattice simulation has not
yet been successfully done due to the sign problem~\cite{Lreview}:
The fermion determinant is not positively definite in the presence of a nonzero baryon
chemical potential $\mu_{\text B}$.

To study the nature of strongly interacting matter at finite density, we thus look for
some special theories that possess a positively definite fermion determinant. One case
is the so-called QCD-like theories at finite baryon density~\cite{QC2D}, where quarks are
in a real or pseudoreal representation of the gauge group, including two-color QCD with
quarks in the fundamental representation and QCD with quarks in the adjoint
representation. While these theories are not real QCD, they can be simulated on the
lattice~\cite{LQC2D} and may give us some information of real QCD at finite baryon density.
Another interesting case is real QCD at finite isospin chemical potential $\mu_{\text I}$
~\cite{ISO}, where the chemical potentials for light $u$ and $d$ quarks have opposite
signs and hence the fermion determinant is positively definite. Chiral perturbation theory
and other effective models predict a continuous quantum phase transition from the vacuum
to the matter phase at $\mu_{\text I}$ equal to the pion mass
$m_\pi$~\cite{ISO,ISOother01,ISOother02}, in contrast to the finite $\mu_{\rm B}$ case,
where the phase transition takes place at $\mu_{\text B}$ approximately equal to the
nucleon mass $m_{\rm N}$. This transition has also been verified by lattice simulations
of QCD at finite isospin chemical potential~\cite{Liso}. The resulting matter near the
quantum phase transition is a dilute Bose condensate of pions with weakly repulsive
interactions~\cite{Bose01}.

The Bose-Einstein condensation (BEC) phenomenon is believed to widely exist in dense
strongly interacting matter. For example, pions or kaons can condense in neutron star matter
if the electron chemical potential exceeds the effective mass for the pions or kaons
~\cite{PiC01,PiC02,PiC03,PiC04}. However, the condensation of pions and kaons in neutron
star matter is rather complicated due to the meson-nucleon interactions in dense nuclear
medium. On the other hand, at asymptotically high density, perturbative QCD calculations
show that the ground state of dense QCD is a weakly coupled BCS superfluid with the
condensation of overlapping Cooper pairs \cite{pQ}. For QCD at finite isospin density,
it is interesting that the dense BCS superfluid and the dilute pion condensate have the
same symmetry-breaking pattern and thus are continued with one another\cite{ISO}. They
are both characterized by the nonzero expectation value $\langle\bar{u}i\gamma_5d\rangle\neq0$.
In condensed matter physics, this phenomenon was first discussed by Eagles~\cite{Eagles}
and Leggett~\cite{Leggett} and is now called BEC-BCS crossover\cite{BCSBEC}.
The BEC-BCS crossover in dense relativistic systems as well as in dense QCD matter has
been extensively studied in recent years~\cite{RBCSBEC,BCSBECQCD}.

While the two-flavor effective models of QCD may be sufficient to describe the phenomenon
of pion condensation at finite isospin density, it is interesting to study the roles of the
strangeness degree of freedom and the U$_{\rm A}(1)$ anomaly. The Nambu--Jona-Lasinio (NJL)
model~\cite{NJL} with quarks as elementary blocks, which describes well the mechanism of
chiral symmetry breaking and low-energy phenomenology of the QCD vacuum, is generally
believed to work at low and moderate temperatures and densities~\cite{NJLreview}.
The pion and kaon condensation in the three-flavor NJL model without U$_{\rm A}(1)$ anomaly
has been studied by Barducci \emph{et al.}~\cite{3NJL01} and by Warringa \emph{et al.}~\cite{3NJL02}.
It is well known that the Kobayashi-Maskawa-'t Hooft (KMT)
term, which mimics the U$_{\rm A}(1)$ anomaly in the NJL model, is crucial to describe the
$\eta-\eta^\prime$ mass splitting and therefore the meson spectra of QCD~\cite{NJLreview}.
Therefore, to study the meson spectra at finite isospin chemical potential, we need to
consider the KMT term.

The Lagrangian density of the three-flavor Nambu--Jona-Lasinio model is given by~\cite{NJLreview}
\begin{eqnarray}
{\cal L}_{\rm{NJL}}=\bar\psi(i\gamma_\mu\partial^\mu-\hat m_0)\psi
+{\cal L}_{\rm S}+{\cal L}_{\rm{KMT}},
\end{eqnarray}
where $\psi=(u,d,s)^{\rm T}$ denotes the quark field and $\hat m_0$ represents the current
quark mass matrix $\hat m_0=\textup{diag}(m_u,m_d,m_s)$. In this work we assume isospin
symmetry, i.e., $m_u=m_d\equiv m_l$. The four-fermion interaction term ${\cal L}_{\rm S}$
is given by
\begin{eqnarray}
{\cal L}_{\rm S}=G\sum_{\alpha=0}^{N_f^2-1}
\left[(\bar\psi\lambda_\alpha\psi)^2
+(\bar\psi i\gamma_5\lambda_\alpha\psi)^2\right],
\label{LFour}
\end{eqnarray}
where $N_f=3$ for our three-flavor model, $\lambda_\alpha$ are the Gell-Mann matrices in
flavor space with $\lambda_0=\sqrt{2/3}{\bf I}$, and $G$ is the coupling constant. This
term represents the interactions with flavor $\rm{U}_{\rm L}(3)\times {\rm U}_{\rm R}(3)$
symmetry in scalar and pseudoscalar channels. However, in real QCD the flavor symmetry
(in the chiral limit $m_l=m_s=0$) is broken down to $\rm{SU}_L(3)\times \rm{SU}_R(3)\times \rm{U}_B(1)$
due to the $\rm{U}_A(1)$ anomaly. In the NJL model, this can be realized by adding the
so-called KMT term ${\cal L}_{\rm{KMT}}$ to the Lagrangian density. It reads
\begin{eqnarray}
{\cal L}_{\rm{KMT}}=-K\left[\textup{det}\bar\psi(1+\gamma_5)\psi
+\textup{det}\bar\psi(1-\gamma_5)\psi\right].\label{LKMT}
\end{eqnarray}
The KMT term contains six-fermion interactions in the three-flavor case. The coupling constant
$G$ and the $\rm{U}_A(1)$ breaking strength $K$ can be determined by the vacuum phenomenology
of QCD. It has been shown that the three-flavor NJL model with KMT term describes well the
meson spectra of QCD, especially the $\eta-\eta^\prime$ mass splitting~\cite{NJLreview}.
In this paper we shall present a systematic study of this model at finite isospin chemical potential $\mu_{\rm I}$.

The above three-flavor NJL model also enables us to study the behavior of the topological susceptibility $\chi$ at
finite isospin chemical potential $\mu_{\rm I}$. The topological susceptibility $\chi$ is a fundamental correlation
function in QCD and is the key to understanding much of the distinctive dynamics in the U$_{\rm A}(1)$ channel. The
general expression of $\chi$ to the leading order in the $1/N_c$ expansion at finite temperature and vanishing
chemical potentials for the three-flavor NJL model has been derived by Fukushima \emph{et al.}~\cite{Fukushima}.
It was shown that the temperature effect suppresses the topological susceptibility $\chi$. On the other hand,
the finite-temperature behavior of the topological susceptibility can also be determined by lattice QCD~\cite{ChiL}.
Therefore, the temperature dependence of the U$_{\rm A}(1)$ anomaly strength $K$ in the NJL model may be determined
by using the lattice data. A relatively small $K$ at large temperature may signal an effective restoration of the
U$_{\rm A}(1)$ anomaly. In this paper we will calculate the topological susceptibility $\chi$ at finite isospin
chemical potential $\mu_{\rm I}$. Since we have no available lattice data for $\chi$ at finite $\mu_{\rm I}$, we
will treat $K$ as a constant. While one may expect a similar suppression of $\chi$ as in the case at finite temperature,
we find instead a clear enhancement at finite isospin density. Our model prediction may be tested by future
lattice simulations, and the $\mu_{\rm I}$ dependence of the U$_{\rm A}(1)$ anomaly
strength $K$ may also be determined by using our expression of $\chi$.

The paper is organized as follows. In Sec. \ref{sec2}, we study the three-flavor NJL model in the mean-field approximation,
which enables us to obtain the chiral and pion condensates, the phase diagram, and the equation of state at finite isospin
chemical potential. In Sec. \ref{sec3}, we construct the effective four-fermion interaction of the KMT term in the presence
of the pion condensation. Using the effective four-fermion interaction, the mesonic excitations at finite isospin chemical
potential are studied within the random-phase approximation. The topological susceptibility at finite isospin chemical
potential is investigated in Sec. \ref{sec4}. We summarize in Sec. \ref{sec5}.

\section{Meson condensation in mean-field approximation}\label{sec2}
To study the three-flavor NJL model at finite chemical potentials and temperature, we introduce the chemical potential matrix
$\hat\mu=\text{diag}(\mu_u,\mu_d,\mu_s)$, where
\begin{eqnarray}
 \mu_u=\frac{\mu_{\rm B}}{3}+\frac{\mu_{\rm I}}{2},\ \ \
 \mu_d=\frac{\mu_{\rm B}}{3}-\frac{\mu_{\rm I}}{2},\ \ \
 \mu_s=\frac{\mu_{\rm B}}{3}-\mu_{\rm S}.
\end{eqnarray}
Here $\mu_{\rm B}$, $\mu_{\rm I}$ and $\mu_{\rm S}$ are referred to as the baryon chemical potential, isospin chemical potential,
and strangeness chemical potential, respectively. At finite temperature and chemical potentials, the partition function of the
NJL model reads
\begin{eqnarray}
Z(T,\mu_{\rm B},\mu_{\rm I},\mu_{\rm S})= \int[d\bar\psi][d\psi]
\exp\left[-\int_0^\beta d\tau \int d^3\bf{x}{\cal L}\right],
\end{eqnarray}
where ${\cal L}={\cal L}_{\rm{NJL}}+\bar{\psi}\hat{\mu}\gamma_0\psi$
and $\beta$ is the inverse of the temperature $T$.

However, the partition function $Z(T,\mu_{\rm B},\mu_{\rm I},\mu_{\rm S})$ cannot be evaluated precisely. The standard approach for
the NJL model is to replace some composite (meson) fields by their expectation values and therefore replace the Lagrangian density
${\cal L}$ by its mean-field approximation ${\cal L}_{\rm{mf}}$~\cite{NJLreview}. To this end, we first
introduce the chiral condensates for the three quark flavors,
\begin{eqnarray}
\sigma_u=\langle\bar uu\rangle,\ \
\sigma_d=\langle\bar dd\rangle,\ \
\sigma_s=\langle\bar ss\rangle.
\end{eqnarray}
It is well known that these condensates are nonzero in the vacuum, corresponding to the spontaneous chiral symmetry breaking of the
QCD vacuum in the chiral limit $m_l=m_s=0$. At finite isospin and strangeness chemical potentials, some pseudoscalar condensates
may arise, corresponding to the condensation of pions and kaons. Therefore, we also introduce the following pseudoscalar condensates:
\begin{eqnarray}
\phi_{ud}=2\langle\bar ui\gamma_5d\rangle,\
\phi_{us}=2\langle\bar ui\gamma_5s\rangle,\
\phi_{ds}=2\langle\bar di\gamma_5s\rangle.
\end{eqnarray}
A nonzero isospin chemical potential $\mu_{\rm I}$ breaks the isospin symmetry $\rm{SU}_{\rm I}(2)$ down to $\rm{U}_{\rm I}(1)$ with generator $I_3$.
The nonzero expectation value of the pseudoscalar condensate $\phi_{ud}$ spontaneously breaks the residual $\rm{U}_{\rm I}(1)$ symmetry, corresponding to Bose-Einstein condensation of charged pions. On the other hand, if $\phi_{us}\neq0$ or $\phi_{ds}\neq0$, the $\rm{U}_{\rm S}(1)$ symmetry corresponding to the conservation of strangeness becomes spontaneously broken, corresponding to the Bose-Einstein condensation of kaons.

The general form of the mean-field Lagrangian density ${\cal L}_{\rm mf}$ takes the form
\begin{eqnarray}
{\cal L}_{\rm mf}=\bar\psi({\cal S}_0^{-1}+\Sigma)\psi-{\cal V}
\end{eqnarray}
where ${\cal S}_0^{-1}=i\gamma_\mu\partial^\mu+\hat{\mu}\gamma_0-\hat m_0$, $\Sigma$ is the quark self-energy in the mean-field approximation, and ${\cal V}$ is the condensation energy that is independent of the quark fields. In the presence of pion and kaon condensation, the self-energy is not diagonal in flavor and can be expressed as
\begin{equation}
\Sigma=\left(\begin{array}{ccc}\Sigma_{uu} & \Sigma_{ud} & \Sigma_{us} \cr
\Sigma_{du} & \Sigma_{dd} & \Sigma_{ds} \cr
\Sigma_{su} & \Sigma_{sd} & \Sigma_{ss} \end{array}\right).
\end{equation}
Accordingly, the quark propagator ${\cal S}$ is also not diagonal. To get the mean-field Lagrangian density ${\cal L}_{\rm mf}$, we replace any composite field operator ${\cal O}$ by ${\cal O}=\langle{\cal O}\rangle+\delta$ and keep only the lowest order in the fluctuation $\delta$. For the KMT term ${\cal L}_{\rm{KMT}}$, it contains six-fermion interaction. The determinant is taken over the flavor space. We have
 \begin{equation}
\det\bar{\psi}\Gamma\psi=\det\left(\begin{array}{ccc}\bar{u}\Gamma u & \bar{u}\Gamma d & \bar{u}\Gamma s \cr
\bar{d}\Gamma u & \bar{d}\Gamma d & \bar{d}\Gamma s \cr
\bar{s}\Gamma u & \bar{s}\Gamma d & \bar{s}\Gamma s \end{array}\right),
\end{equation}
 where $\Gamma=1\pm\gamma_5$. Therefore, in the presence of meson condensates $\phi_{ud}$, $\phi_{us}$, and $\phi_{ds}$, the off-diagonal components contribute and the mean-field approximation of the KMT term should be taken carefully.

After some tedious calculations, we obtain the expression for the condensation energy,
 \begin{eqnarray}
{\cal V}&=&2G(\sigma_u^2+\sigma_d^2+\sigma_s^2)
+G(|\phi_{ud}|^2+|\phi_{us}|^2+|\phi_{ds}|^2)\nonumber\\
&-&4K\sigma_u\sigma_d\sigma_s
-K(|\phi_{ud}|^2\sigma_s+|\phi_{us}|^2\sigma_d+|\phi_{ds}|^2\sigma_u).
\end{eqnarray}
With the result of the self-energy, the inverse of the quark propagator ${\cal S}$ can be expressed as
\begin{eqnarray}
 &&{\cal S}^{-1}={\cal S}_0^{-1}+\Sigma\nonumber\\
&=&\left(\begin{array}{ccc}i\gamma^\mu\partial_\mu-M_u & \Sigma_{ud} & \Sigma_{us} \cr
 \Sigma_{du} &i\gamma^\mu\partial_\mu-M_d & \Sigma_{ds} \cr
 \Sigma_{su}&\Sigma_{sd} & i\gamma^\mu\partial_\mu-M_s\end{array}\right)+\hat{\mu}\gamma_0,
\end{eqnarray}
where the effective quark masses are defined as
\begin{eqnarray}
M_u&=&m_{u}-4G\sigma_u+2K\sigma_d\sigma_s+\frac{K}{2}|\phi_{ds}|^2,\nonumber\\
M_d&=&m_{d}-4G\sigma_d+2K\sigma_u\sigma_s+\frac{K}{2}|\phi_{us}|^2,\nonumber\\
M_s&=&m_{s}-4G\sigma_s+2K\sigma_u\sigma_d+\frac{K}{2}|\phi_{ud}|^2
\end{eqnarray}
and the off-diagonal components of the self-energy read
\begin{eqnarray}
&&\Sigma_{ud}=\Sigma_{du}^*=i\gamma_5(2G-K\sigma_s)\phi_{ud}^*+\frac{K}{2}\phi_{us}^*\phi_{ds},\nonumber\\
&&\Sigma_{us}=\Sigma_{su}^*=i\gamma_5(2G-K\sigma_d)\phi_{us}^*+\frac{K}{2}\phi_{ud}^*\phi_{ds}^*,\nonumber\\
&&\Sigma_{ds}=\Sigma_{sd}^*=i\gamma_5(2G-K\sigma_u)\phi_{ds}^*+\frac{K}{2}\phi_{ud}\phi_{us}^*.
\end{eqnarray}

An alternative way for calculating the mean-field approximation to the self-energy $\Sigma$ is to Wick-contract all
possible pairs of creation and destruction field operators in the Lagrangian, so that only two remain ``alive" as
field operators. To the lowest order in the $1/N_c$ expansion, the result from this method should be consistent with
the mean-field approximation. In the absence of meson condensates, the above expressions recover the well-known result
for the three-flavor NJL model~\cite{NJLreview}. However, in the presence of meson condensates, we find that they
contribute to the effective quark masses $M_i(i=u,d,s)$ due to the ${\rm U}_{\rm A}(1)$ anomaly.

Having obtained the mean-field approximation of the Lagrangian density, we can work out the path integral by replacing
${\cal L}$ with ${\cal L}_{\rm mf}$. The mean-field approximation to the thermodynamic potential reads
\begin{eqnarray}
\Omega_{\rm mf}={\cal V}-\frac{1}{\beta V}\rm{Tr}\ln {\cal S}^{-1}.
\end{eqnarray}
In general, the physical values of the condensates can be obtained by minimizing the thermodynamic potential at given temperature
and chemical potentials. In this work, we focus on the case $\mu_{\rm B}=\mu_{\rm S}=0$ and $\mu_{\rm I}\neq0$. In this case, we have
$\sigma_u=\sigma_d\equiv\sigma_l$ and $\phi_{us}=\phi_{ds}=0$ (no kaon condensation). The thermodynamic potential can be analytically evaluated
as
\begin{eqnarray}
\Omega_{\rm mf}&=&G(4\sigma_l^2+2\sigma_s^2+|\phi_{ud}|^2)-K\sigma_s(4\sigma_l^2+|\phi_{ud}|^2)\nonumber\\
&-&2N_c\int\frac{d^3{\bf k}}{(2\pi)^3}\left(E_{\bf k}^{+}+E_{\bf k}^{-}+E_{\bf k}^s\right)\nonumber\\
&-&4N_cT\int\frac{d^3{\bf k}}{(2\pi)^3}\bigg[\ln\left(1+e^{-\beta E_{\bf k}^{+}}\right)\nonumber\\
&&+\ln\left(1+e^{-\beta E_{\bf k}^{-}}\right)+\ln\left(1+e^{-\beta E_{\bf k}^s}\right)\bigg]\label{ThermalP},
\end{eqnarray}
where the dispersions are defined as
\begin{eqnarray}
E_{\bf k}^\pm&=&\sqrt{(E_{\bf k}^l\pm\mu_{\rm I}/2)^2+|\Delta|^2},\nonumber\\
E_{\bf k}^l&=&\sqrt{{\bf k}^2+M_l^2},\nonumber\\
E_{\bf k}^s&=&\sqrt{{\bf k}^2+M_s^2}.
\end{eqnarray}
Here the effective quark masses for light and strange quarks have been reduced to
\begin{eqnarray}
M_l&=&m_{q}-4G\sigma_l+2K\sigma_l\sigma_s,\nonumber\\
M_s&=&m_{s}-4G\sigma_s+2K\sigma_l^2+\frac{K}{2}|\phi_{ud}|^2,
\end{eqnarray}
and the quantity $\Delta$ is given by
\begin{eqnarray}
\Delta=(2G-K\sigma_s)\phi_{ud}.
\end{eqnarray}
Since the thermodynamic potential depends only on $|\phi_{ud}|^2$, we can set $\phi_{ud}$ and $\Delta$ to be real values without loss of
generality. This is related to the spontaneous breaking of the U$_{\rm I}(1)$ symmetry, which means the phase of the order parameter
$\phi_{ud}$ can be chosen arbitrarily.

In the momentum space, the quark propagator ${\cal S}(k)$ takes the form
\begin{equation}
{\cal S}(k)=\left(\begin{array}{ccc}{\cal S}_{uu}(k) & {\cal S}_{ud}(k) & 0 \cr
{\cal S}_{du}(k) & {\cal S}_{dd}(k) & 0 \cr 0 & 0 & {\cal S}_{s}(k) \end{array}\right), \label{Qpropagator}
\end{equation}
where $k=(i\omega_n,{\bf k})$ with $\omega_n=(2n+1)\pi T$ ($n$ integer) being the fermion Matsubara frequency. The nonzero components can be analytically evaluated as~\cite{massive}
\begin{eqnarray}
{\cal S}_{uu}(k)&=&\frac{i\omega_n+\xi_{\bf k}^-}{(i\omega_n)^2-(E_{\bf k}^-)^2}\Lambda_+^l\gamma_0
+\frac{i\omega_n-\xi_{\bf k}^+}{(i\omega_n)^2-(E_{\bf k}^+)^2}\Lambda_-^l\gamma_0,\nonumber\\
{\cal S}_{dd}(k)&=&\frac{i\omega_n-\xi_{\bf k}^-}{(i\omega_n)^2-(E_{\bf k}^-)^2}\Lambda_-^l\gamma_0
+\frac{i\omega_n+\xi_{\bf k}^+}{(i\omega_n)^2-(E_{\bf k}^+)^2}\Lambda_+^l\gamma_0,\nonumber\\
{\cal S}_{ud}(k)&=&\frac{i\Delta}{(i\omega_n)^2-(E_{\bf k}^-)^2}\Lambda_+^l\gamma_5
+\frac{i\Delta}{(i\omega_n)^2-(E_{\bf k}^+)^2}\Lambda_-^l\gamma_5,\nonumber\\
{\cal S}_{du}(k)&=&\frac{i\Delta}{(i\omega_n)^2-(E_{\bf k}^-)^2}\Lambda_-^l\gamma_5
+\frac{i\Delta}{(i\omega_n)^2-(E_{\bf k}^+)^2}\Lambda_+^l\gamma_5,\nonumber\\
{\cal S}_{s}(k)&=&\frac{1}{i\omega_n-E_{\bf k}^s}\Lambda_+^s\gamma_0
+\frac{1}{i\omega_n+E_{\bf k}^s}\Lambda_-^s\gamma_0,
\end{eqnarray}
where $\xi_{\bf k}^\pm=E_{\bf k}^l\pm\mu_{\rm I}/2$ and the energy projectors $\Lambda_\pm^l$ and $\Lambda_\pm^s$ are defined as
\begin{eqnarray}
\Lambda_\pm^l&=&\frac{1}{2}\left[1\pm\frac{\gamma_0(\mbox{\boldmath{$\gamma$}}\cdot{\bf k}+M_l)}{E_{\bf k}^l}\right],\nonumber\\
\Lambda_\pm^s&=&\frac{1}{2}\left[1\pm\frac{\gamma_0(\mbox{\boldmath{$\gamma$}}\cdot{\bf k}+M_s)}{E_{\bf k}^s}\right].
\end{eqnarray}

The gap equations that determine the physical values of the chiral and pion condensates can be obtained by minimizing the thermodynamic
potential $\Omega_{\rm mf}$. Self-consistently, they can also be derived from the Green's function relations. We have
\begin{eqnarray}
\sigma_l&=&\frac{N_c}{\beta}\sum_n\int\frac{d^3{\bf k}}{(2\pi)^3}{\rm Tr}_{\rm D}{\cal S}_{uu}(k)\nonumber\\
&=&\frac{N_c}{\beta}\sum_n\int\frac{d^3{\bf k}}{(2\pi)^3}{\rm Tr}_{\rm D}{\cal S}_{dd}(k),\nonumber\\
\sigma_s&=&\frac{N_c}{\beta}\sum_n\int\frac{d^3{\bf k}}{(2\pi)^3}{\rm Tr}_{\rm D}{\cal S}_{s}(k),\nonumber\\
\phi_{ud}&=&2\frac{N_c}{\beta}\sum_n\int\frac{d^3{\bf k}}{(2\pi)^3}{\rm Tr}_{\rm D}\left[{\cal S}_{du}(k)i\gamma_5\right],
\end{eqnarray}
where the trace runs only in the spin space. Finally, we get the explicit expressions of the gap equations
\begin{eqnarray}
&&\sigma_s=-2N_cM_s\int\frac{d^3{\bf k}}{(2\pi)^3}\frac{1-2f(E_{\bf k}^s)}{E_{\bf k}^s},\nonumber\\
&&\sigma_l=-N_cM_l\int\frac{d^3{\bf k}}{(2\pi)^3}\frac{1}{E_{\bf k}^l}
\sum_{\alpha=\pm}\frac{\xi_{\bf k}^\alpha}{E_{\bf k}^\alpha}(1-2f(E_{\bf k}^\alpha)),\nonumber\\
&&\phi_{ud}=2N_c\Delta\int\frac{d^3{\bf k}}{(2\pi)^3}
\sum_{\alpha=\pm}\frac{1-2f(E_{\bf k}^\alpha)}{E_{\bf k}^\alpha}, \label{gapeq}
\end{eqnarray}
where $f(E)=1/(e^{\beta E}+1)$ is the Fermi-Dirac distribution.

Now we turn to numerical results. The NJL model is nonrenormalizable and we introduce an ultraviolet cutoff $\Lambda$ for the
three-momentum to regularize the divergent integrals. The model therefore has five parameters, the current quark masses $m_l$ and
$m_s$, the coupling constant $G$, the anomaly coupling $K$, and the cutoff $\Lambda$. For the light quark mass, we choose
$m_l=5.5$MeV. The others can be determined by recovering the well-measured vacuum phenomenology. Here we employ the pion decay
constant and the masses of the pion, kaon, and the $\eta^\prime$ meson. The data we adopted are $f_\pi=92.4$,
$m_\pi=135.0$, $m_K=497.7$, and $m_{\eta'}=957.8\rm{MeV}$. The model parameters obtained by fitting these data are
$m_l=5.5$, $m_s=140.7$, $\Lambda=602.3$MeV, $G\Lambda^2=1.835$, and $K\Lambda^5=12.36$~\cite{Renberg}. Such a parameter set leads
to the quark chiral condensate $\langle\bar{l}l\rangle_0=-(241.9{\rm MeV})^3$ and $\langle\bar{s}s\rangle_0=-(257.7{\rm MeV})^3$ and
the constituent quark masses $M_l=367.7$ and $M_s=549.5$MeV.

\subsection{Condensates and phase diagram}

\begin{figure}[hbt]
\centering
\includegraphics[width=7.5cm]{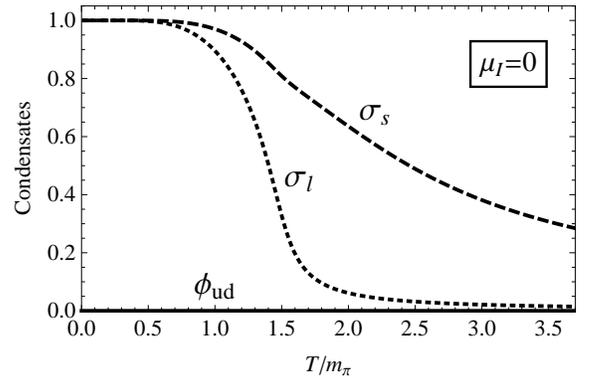}
\caption{The chiral and pion condensates as functions of the temperature $T$
at vanishing isospin chemical potential $\mu_{\rm I}=0$. $\sigma_l$ and
$\phi_{ud}$ are scaled by $\langle\bar{l}l\rangle_0$ and
2$\langle\bar{l}l\rangle_0$, respectively, and $\sigma_s$ is scaled by
$\langle\bar{s}s\rangle_0$. In the following we use the same quantities to
scale the chiral and pion condensates.} \label{condensate01}
\end{figure}

In Fig. \ref{condensate01} we show the behavior of the chiral and pion condensates
with increasing temperature $T$ at vanishing isospin chemical potential. In this
case, pion condensation does not arise. The system undergoes a crossover from the
low-temperature hadronic matter to the high-temperature quark matter. The results
for the chiral condensates shown here are consistent with those reported in the
previous studies~\cite{Renberg,Costa}.

\begin{figure}[hbt]
\centering
\includegraphics[width=7.5cm]{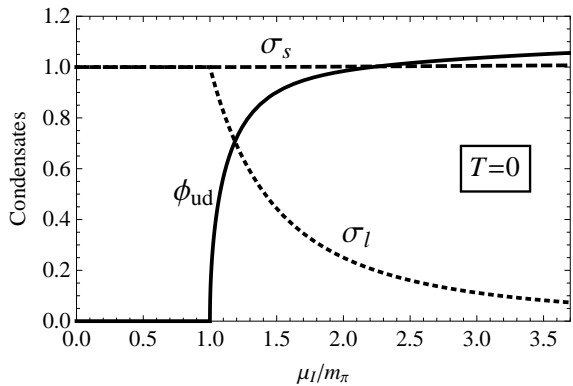}
\caption{The chiral and pion condensates as functions of the isospin chemical
potential $\mu_{\rm I}$ at zero temperature.} \label{condensate02}
\end{figure}

We are most interested in the case of zero temperature and finite isospin chemical potential.
Since the charged pions ($\pi^\pm$) are the lightest mesons that carry isospin quantum number,
we expect that they get condensed as long as the isospin chemical potential $\mu_{\rm I}$ exceeds
a critical value. Without loss of generality, we study the case $\mu_{\rm I}>0$. Solving the
coupled gap equations, we obtain the evolution of the chiral and pion condensates with
increasing isospin chemical potential, as shown in Fig. \ref{condensate02}. We find that the
onset of pion condensation is located precisely at $\mu_{\rm I}=m_\pi$, as has been shown
analytically in the two-flavor NJL model~\cite{ISOother02}. In the three-flavor model, this
can also be shown explicitly. In the regime $|\mu_{\rm I}|<m_\pi$, the chiral condensates keep
their values in the vacuum, which means that the system stays in the vacuum for $|\mu_{\rm I}|<m_\pi$
and no isospin charge is excited. At the onset of pion condensation, we have from the gap equations
\begin{eqnarray}
1=2N_c(2G-K\sigma_s)\int\frac{d^3{\bf k}}{(2\pi)^3}
\frac{2E_{\bf k}^l}{(E_{\bf k}^l)^2-\mu_{\rm I}^2/4}.
\end{eqnarray}
This equation coincides with the mass equation for the pion in the vacuum and gives precisely
$\mu_{\rm I}=m_\pi$.

\begin{figure}[hbt]
\centering
\includegraphics[width=7.5cm]{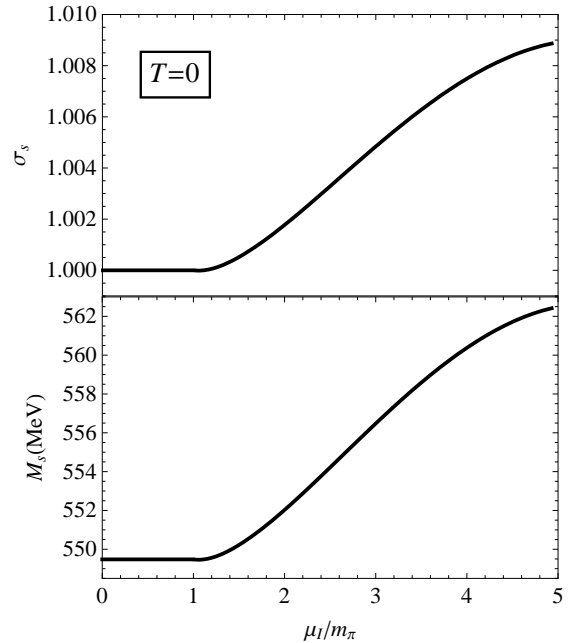}
\caption{A zoom look at the behavior of the strange quark chiral condensate
$\sigma_s$ and the strange quark effective mass $M_s$ at finite isospin
chemical potential.} \label{condensate03}
\end{figure}

When the isospin chemical potential $\mu_{\rm I}$ exceeds the critical value $m_\pi$, the
light quark chiral condensate $\sigma_l$ decreases significantly and the isospin density is
generated due to the appearance of condensed pions. The behavior of the pion condensate
$\phi_{ud}$ and the light quark chiral condensate $\sigma_l$ is similar to that found in the
two-flavor model~\cite{ISOother02}. Let us now focus on the roles of the strangeness and the
${\rm U}_{\rm A}(1)$ anomaly. In the absence of the ${\rm U}_{\rm A}(1)$ anomaly, i.e., $K=0$, the
gap equation for the strange quark chiral condensate $\sigma_s$ decouples from the other two
equations. Therefore, the strangeness degree of freedom has nothing to do with the pion
condensation for $K=0$. The strange quark chiral condensate $\sigma_s$ just keeps its vacuum
value no matter how large $\mu_{\rm I}$ is. However, in the presence of the ${\rm U}_{\rm A}(1)$
anomaly, the strange quark chiral condensate $\sigma_s$ couples to the other condensates
through the nonvanishing anomaly strength $K$. Therefore, the strange quark chiral condensate
should change in the pion condensed phase. However, from its behavior in Fig. \ref{condensate02},
we find that its change is very slight. In Fig. \ref{condensate03}, we show a zoom plot for the
strange quark chiral condensate $\sigma_s$. It starts to increase at the onset of pion condensation.
However, the value is only increased by about $1\%$ at $\mu_{\rm I}=5m_\pi$. The behavior of the
strange quark effective mass $M_s$ is also shown in Fig. \ref{condensate03}. It is increased by
about $12$MeV at $\mu_{\rm I}=5m_\pi$. Therefore, our studies of the three-flavor NJL model verify
the validity of the two-flavor models in the description of pion condensation.

\begin{figure}[hbt]
\centering
\includegraphics[width=7.5cm]{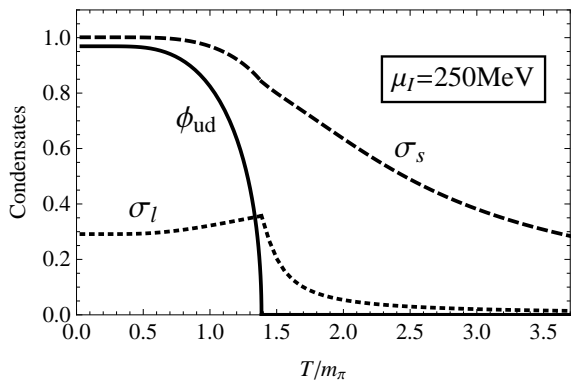}
\caption{The chiral and pion condensates as functions of the temperature $T$
at fixed isospin chemical potential $\mu_{\rm I}=250$MeV.} \label{condensate04}
\end{figure}

At finite temperature, the pion condensate gets melted. In general there exists a phase transition
from the pion superfluid phase to the normal phase with $\phi_{ud}=0$. In the present model, we find
this transition is of second order. In Fig. \ref{condensate04}, we show the finite-temperature
behavior of the chiral and pion condensates at $\mu_{\rm I}>m_\pi$. The temperature effect, while
reduces the pion condensate $\phi_{ud}$ and the strange quark chiral condensate $\sigma_s$, enhances
the light quark chiral condensate $\sigma_q$ in the superfluid domain $T<T_c$. At high temperature,
all condensates are suppressed by the temperature effect.

\begin{figure}[hbt]
\centering
\includegraphics[width=7.5cm]{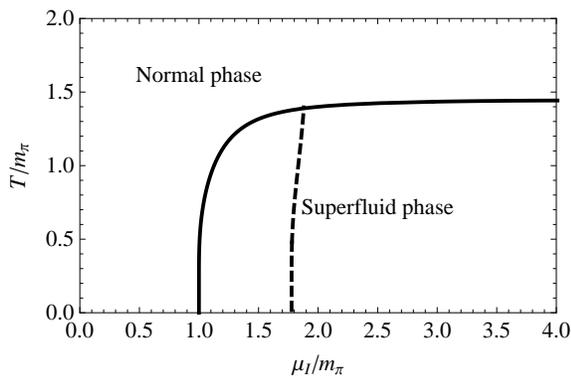}
\caption{Phase diagram of the three-flavor NJL model in the $T$-$\mu_{\rm I}$
plane. The normal and superfluid phases correspond to the phases with vanishing
and nonvanishing pion condensates, respectively. The black dashed line denotes
the BEC-BCS crossover from pion to quark-antiquark condensation.}
\label{phasediagram}
\end{figure}

In Fig. \ref{phasediagram}, we show the phase diagram of
the present model in the $T$-$\mu_{\rm I}$ plane. The phase boundary between the superfluid phase
and the normal phase is determined by the line where the pion condensate $\phi_{ud}$ vanishes. For
the normal phase in the domain $\mu_{\rm I}<m_\pi$, there exists a crossover from the low-temperature
hadronic matter to the high-temperature quark matter that we do not show explicitly. In the superfluid
phase we also expect a BEC-BCS crossover from pion condensation to quark-antiquark condensation.
Near the onset of pion condensation $\mu_{\rm I}=m_\pi$, the system can be identified as a dilute Bose-Einstein
condensate of weakly interacting pions. However, at large isospin density, the pions no longer remain
tightly bound bosons, and the system becomes a BCS superfluid with condensation of quark-antiquark pairs
(but with the same quantum number of pions). According to the spirit of the BEC-BCS crossover~\cite{BCSBEC},
the crossover can be estimated by studying the fermionic excitation spectrum. Here we focus on the excitation spectrum
$E_{\bf k}^-$~\cite{RBCSBEC,BCSBECQCD}. Its minimum remains at ${\bf k}=0$ for $M_l>\mu_{\rm I}/2$ but shifts to nonzero momentum if
$M_l<\mu_{\rm I}/2$. The BEC-BCS crossover line can be estimated by the condition $M_l(\mu_{\rm I})=\mu_{\rm I}/2$.
This line is shown in Fig. \ref{phasediagram}. A clear picture the BEC-BCS crossover can be shown by studying
the behavior of the pions above the superfluid critical temperature $T_c$. In the BEC domain, the pions are bound
bosons above $T_c$. However, they become loose resonances at the BCS side.

\subsection{Equation of state}

In the final part of the section, we study the equation of state of the isospin matter.
Here we focus on the zero temperature case.

\begin{figure}[hbt]
\centering
\includegraphics[width=7.5cm]{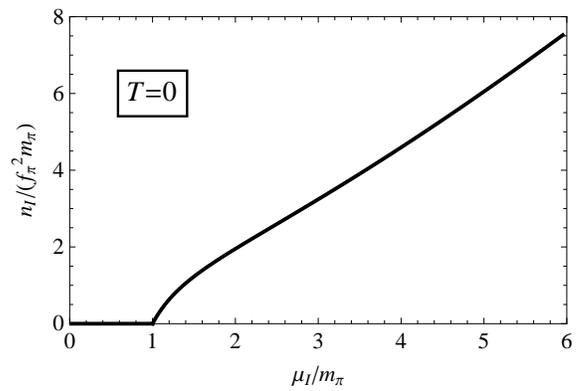}
\caption{The isospin density $n_{\rm I}$ (scaled by $f_\pi^2m_\pi$) as a function
of the isospin chemical potential $\mu_{\rm I}$.} \label{isospindensity}
\end{figure}

\begin{figure}[hbt]
\centering
 \includegraphics[width=7.5cm]{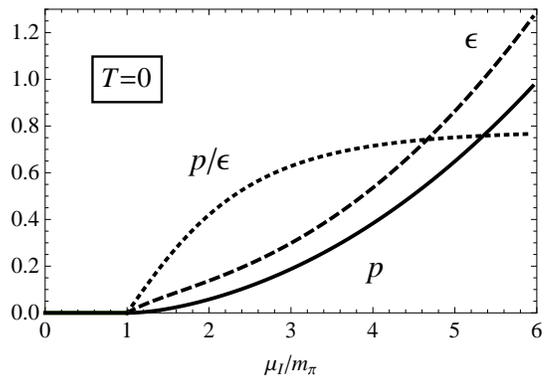}
\caption{The pressure and energy densities, $p$ and $\epsilon$, as functions
of the isospin chemical potential $\mu_{\rm I}$. They have been scaled by the
quantity $20f_\pi^2m_\pi^2$. The ratio $p/\epsilon$ is also shown.} \label{EOS}
\end{figure}

In Fig. \ref{isospindensity}, we show the isospin density $n_{\rm I}$ as a function of
the isospin chemical potential $\mu_{\rm I}$. In the mean-field approximation, the isospin
density $n_{\rm I}$ reads
\begin{eqnarray}
n_{\rm I}=N_c\int\frac{d^3{\bf k}}{(2\pi)^3}\left(\frac{\xi_{\bf k}^+}{E_{\bf k}^+}-\frac{\xi_{\bf k}^-}{E_{\bf k}^-}\right).
\end{eqnarray}
From this expression, we see clearly that the isospin density remains zero for the regime $|\mu_{\rm I}|<m_\pi$, and it becomes
nonzero in the superfluid phase. Near the onset of pion condensation, the numerical result agrees well with that predicted
by the chiral perturbation theory,
\begin{eqnarray}
n_{\rm I}\simeq f_\pi^2\mu_{\rm I}\left(1-\frac{m_\pi^4}{\mu_{\rm I}^4}\right).
\end{eqnarray}
However, at large isospin chemical potential, this behavior no longer holds, since the system undergoes BEC-BCS crossover and
is no longer Bose-Einstein condensate of weakly interacting pions.

\begin{figure}[hbt]
\centering
\includegraphics[width=9cm]{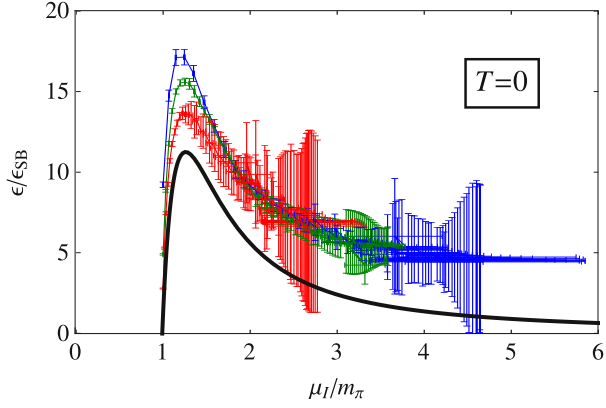}
\caption{(color online). The energy density $\epsilon$ scaled by its Stefan-Boltzmann limit
as a function of the isospin chemical potential $\mu_{\rm I}$. The colorful
data are taken from recent lattice simulation~\cite{EOSiso}. Different colors
correspond to different lattice spatial spacings used in the simulation. } \label{EOSL}
\end{figure}

In Fig. \ref{EOS} the pressure $p$ and the energy density $\epsilon$ are shown as functions of the isospin
chemical potential. They keep vanishing in the regime $|\mu_{\rm I}|<m_\pi$, and arise in the superfluid phase due to the
appearance of condensed pions. In Fig. \ref{EOSL}, we show the energy density $\epsilon$ scaled by its Stefan-Boltzmann limit,
\begin{eqnarray}
\epsilon_{\rm SB}=\frac{N_fN_c}{4\pi^2}\left(\frac{\mu_{\rm I}}{2}\right)^4.
\end{eqnarray}
The result from recent lattice simulation~\cite{EOSiso} with a larger pion mass $m_\pi\sim390$MeV is also shown as a comparison.
We find that the scaled energy density $\epsilon/\epsilon_{\rm SB}$ shows a peak near the onset of the pion condensation
(at $\mu_{\rm I}\simeq 1.3m_\pi$), which is consistent with the lattice result.

\section{Mesonic excitations in the Random-Phase Approximation}\label{sec3}
In the spirit of the Nambu--Jona-Lasinio model, the mesons are regarded as collective excitations~\cite{NJL,NJLreview}.
The meson propagator, to the leading order in $1/N_c$, can be constructed by an infinite sum of the quark-antiquark ring
diagrams associated with the four-fermion interaction, or the so-called random-phase approximation that is recognized to be
a geometric progression. However, in the present three-flavor model, the interaction terms contains not only four-fermion
interaction but also six-fermion interaction (due to the KMT term). Therefore, to study the mesonic excitations at finite
isospin chemical potential and temperature, we should first construct an effective Lagrangian with only effective four-fermion
interactions.

Let us now construct an effective four-fermion interaction ${\cal L}_{\rm KMT}^{4\rm f}$ for the KMT term. We expect that we
obtain the same self-energy $\Sigma$ after taking the mean-field approximation for ${\cal L}_{\rm KMT}^{4\rm f}$. We do this
by contracting out one quark and one antiquark field operator, such that an effective four-fermion interaction is remaining.
Unlike the procedure for the vacuum case where $\phi_{ud}=0$, here we should take care of the effect of the pion condensate
$\phi_{ud}$ that leads to scalar-pseudoscalar mixing. After a tedious calculation, we obtain
\begin{eqnarray}
{\cal L}_{\rm KMT}^{4\rm f}&=&{\cal L}_{\rm KMT}^{\rm s}+{\cal L}_{\rm KMT}^{\rm p}+{\cal L}_{\rm KMT}^{\rm sp},
\end{eqnarray}
where ${\cal L}_{\rm KMT}^{\rm s}$ is the scalar interaction part,
\begin{eqnarray}
{\cal L}_{\rm KMT}^{\rm s}&=&-\frac{K}{3}(2\sigma_l+\sigma_s)(\bar\psi\lambda_0\psi)^2+\frac{K}{2}\sigma_s(\bar\psi\lambda_3\psi)^2\nonumber\\
&+&\frac{K}{6}(4\sigma_l-\sigma_s)(\bar\psi\lambda_8\psi)^2\nonumber\\
&+&\frac{\sqrt{2}}{3}K(\sigma_l-\sigma_s)(\bar\psi\lambda_0\psi)(\bar\psi\lambda_8\psi)\nonumber\\
&+&K\sigma_s(\bar\psi\lambda_1^-\psi)(\bar\psi\lambda_1^+\psi)\nonumber\\
&+&K\sigma_l(\bar\psi\lambda_4^-\psi)(\bar\psi\lambda_4^+\psi)\nonumber\\
&+&K\sigma_l(\bar\psi\lambda_6^-\psi)(\bar\psi\lambda_6^+\psi),
\end{eqnarray}
${\cal L}_{\rm KMT}^{\rm p}$ is the pseudoscalar interaction part,
\begin{eqnarray}
{\cal L}_{\rm KMT}^{\rm p}&=&\frac{K}{3}(2\sigma_l+\sigma_s)(\bar\psi i\gamma_5\lambda_0\psi)^2
-\frac{K}{2}\sigma_s(\bar\psi i\gamma_5\lambda_3\psi)^2\nonumber\\
&-&\frac{K}{6}(4\sigma_l-\sigma_s)(\bar\psi i\gamma_5\lambda_8\psi)^2\nonumber\\
&-&\frac{\sqrt{2}}{3}K(\sigma_l-\sigma_s)(\bar\psi i\gamma_5\lambda_0\psi)(\bar\psi i\gamma_5\lambda_8\psi)\nonumber\\
&-&K\sigma_s(\bar\psi i\gamma_5\lambda_1^-\psi)(\bar\psi i\gamma_5\lambda_1^+\psi)\nonumber\\
&-&K\sigma_l(\bar\psi i\gamma_5\lambda_4^-\psi)(\bar\psi i\gamma_5\lambda_4^+\psi)\nonumber\\
&-&K\sigma_l(\bar\psi i\gamma_5\lambda_6^-\psi)(\bar\psi i\gamma_5\lambda_6^+\psi),
\end{eqnarray}
and ${\cal L}_{\rm KMT}^{\rm sp}$ is the scalar-pseudoscalar interaction part,
\begin{eqnarray}
{\cal L}_{\rm KMT}^{\rm sp}&=&
-\frac{\sqrt{3}}{6}K\phi_{ud}^*\left[(\bar\psi i\gamma_5\lambda_0\psi)(\bar\psi\lambda_1^+\psi)
+(\bar\psi\lambda_0\psi)(\bar\psi i\gamma_5\lambda_1^+\psi)\right]\nonumber\\
&-&\frac{\sqrt{3}}{6}K\phi_{ud}\left[(\bar\psi i\gamma_5\lambda_0\psi)(\bar\psi\lambda_1^-\psi)
+(\bar\psi\lambda_0\psi)(\bar\psi i\gamma_5\lambda_1^-\psi)\right]\nonumber\\
&+&\frac{K}{\sqrt{6}}\phi_{ud}^*\left[(\bar\psi i\gamma_5\lambda_1^+\psi)(\bar\psi\lambda_8\psi)
+(\bar\psi\lambda_1^+\psi)(\bar\psi i\gamma_5\lambda_8\psi)\right]\nonumber\\
&+&\frac{K}{\sqrt{6}}\phi_{ud}\left[(\bar\psi i\gamma_5\lambda_1^-\psi)(\bar\psi\lambda_8\psi)
+(\bar\psi\lambda_1^-\psi)(\bar\psi i\gamma_5\lambda_8\psi)\right]\nonumber\\
&+&\frac{K}{2}\phi_{ud}^*\left[(\bar\psi\lambda_6^-\psi)(\bar\psi i\gamma_5\lambda_4^+\psi)
+(\bar\psi i\gamma_5\lambda_6^-\psi)(\bar\psi\lambda_4^+\psi)\right]\nonumber\\
&+&\frac{K}{2}\phi_{ud}\left[(\bar\psi\lambda_6^+\psi)(\bar\psi i\gamma_5\lambda_4^-\psi)
+(\bar\psi i\gamma_5\lambda_6^+\psi)(\bar\psi\lambda_4^-\psi)\right].
\end{eqnarray}
Here we have defined
\begin{eqnarray}
\lambda_1^\pm&=&\frac{1}{\sqrt{2}}\left(\lambda_1\pm i\lambda_2\right),\nonumber\\
\lambda_4^\pm&=&\frac{1}{\sqrt{2}}\left(\lambda_4\pm i\lambda_5\right),\nonumber\\
\lambda_6^\pm&=&\frac{1}{\sqrt{2}}\left(\lambda_6\pm i\lambda_7\right).
\end{eqnarray}
In the absence of pion condensate, $\phi_{ud}=0$, the effective four-fermion interaction recovers the result obtained in the
previous literature~\cite{NJLreview}. The scalar-pseudoscalar interaction ${\cal L}_{\rm KMT}^{\rm sp}$
is purely induced by the pion condensation.

Then the mesonic excitations can be studied by using the effective Lagrangian with only four-fermion interactions,
\begin{eqnarray}
{\cal L}_{\rm eff}=\bar\psi(i\gamma^\mu\partial_\mu-\hat{m}_0)\psi+{\cal L}_{\rm S}+{\cal L}_{\rm KMT}^{4\rm f}.
\end{eqnarray}
For convenience, we rewrite the effective four-fermion interactions in a compact form,
\begin{eqnarray}
{\cal L}_{\rm eff}=\bar\psi(i\gamma^\mu\partial_\mu-\hat m_0)\psi
+{\bf\Phi}^\dagger {\bf G}{\bf \Phi},
\end{eqnarray}
where the bilinear field ${\bf \Phi}$ is defined as
\begin{eqnarray}
{\bf \Phi}=\left(
\begin{array}{c} \bar{\psi}\lambda_1^-\psi \cr \bar{\psi}\lambda_1^+\psi\cr
\bar{\psi}\lambda_3\psi \cr \bar{\psi}\lambda_4^-\psi \cr \bar{\psi}\lambda_4^+\psi\cr
\bar{\psi}\lambda_6^-\psi \cr \bar{\psi}\lambda_6^+\psi \cr
\bar{\psi}\lambda_0\psi \cr \bar{\psi}\lambda_8\psi \cr
\bar{\psi}i\gamma_5\lambda_1^-\psi \cr \bar{\psi}i\gamma_5\lambda_1^+\psi\cr
\bar{\psi}i\gamma_5\lambda_3\psi \cr \bar{\psi}i\gamma_5\lambda_4^-\psi \cr \bar{\psi}i\gamma_5\lambda_4^+\psi\cr
\bar{\psi}i\gamma_5\lambda_6^-\psi \cr \bar{\psi}i\gamma_5\lambda_6^+\psi \cr
\bar{\psi}i\gamma_5\lambda_0\psi \cr \bar{\psi}i\gamma_5\lambda_8\psi
\end{array}\right)
\end{eqnarray}
and the effective coupling matrix ${\bf G}$ takes the block form
\begin{eqnarray}
{\bf G}=\left( \begin{array}{ccccccc} G_{11}&0 &0&0&0&0 & G_{17}\cr
0&G_{22}&0&0&0&0&0\cr 0&0& G_{33}&0&0&G_{36}&0\cr 0&0&0&G_{44}&0&0&0\cr 0&0&0&0 &G_{55}&0 &0
\cr 0&0&G_{63}&0 &0&G_{66} & 0\cr G_{71}&0&0&0 &0&0 & G_{77}
\end{array}\right).
\end{eqnarray}
The nonvanishing blocks of ${\bf G}$ are given by
\begin{eqnarray}
G_{11}&=&\left(G+\frac{K}{2}\sigma_s\right) I_2,\nonumber\\
G_{22}&=&G+\frac{K}{2}\sigma_s,\nonumber\\
G_{33}&=&\left(G+\frac{K}{2}\sigma_l\right) I_4,\nonumber\\
G_{44}&=&GI_4+K\left(\begin{array}{cc}A&\phi_{ud}B^{\rm T} \cr \phi_{ud}B&-\frac{1}{2}\sigma_sI_2\end{array}\right),\nonumber\\
G_{55}&=&G-\frac{K}{2}\sigma_s,\nonumber\\
G_{66}&=&\left(G-\frac{K}{2}\sigma_l\right) I_4,\nonumber\\
G_{77}&=&GI_2-KA, \nonumber\\
G_{17}&=&G_{71}^{\rm{T}}=K\phi_{ud}B, \nonumber\\
G_{36}&=&G_{63}^{\rm{T}}=K\phi_{ud}C,
\end{eqnarray}
where $I_n$ is the $n\times n$ identity matrix, and the matrices $A,B$, and $C$ are defined as
\begin{eqnarray}
A&=&\left(\begin{array}{cc}-\frac{1}{3}(2\sigma_l+\sigma_s)&\frac{\sqrt{2}}{6}(\sigma_l-\sigma_s) \cr
 \frac{\sqrt{2}}{6}(\sigma_l-\sigma_s)&\frac{1}{6}(4\sigma_l-\sigma_s)\end{array}\right),\nonumber\\
B&=&\frac{\sqrt{3}}{12}\left(\begin{array}{cc}-1&\sqrt{2} \cr -1&\sqrt{2}\end{array}\right),\nonumber\\
C&=&\frac{1}{4}\left(\begin{array}{cccc} 0&0&1&0\cr 0&0&0&1 \cr 1&0&0&0 \cr 0&1&0&0\end{array}\right).
\end{eqnarray}

Using the compact form of the effective four-fermion interaction, the meson propagator ${\bf D}(q)$ in the random-phase
approximation can be expressed as
\begin{eqnarray}
{\bf D}(q)=\frac{2{\bf G}}{{\bf 1}-2{{\bf G}{\bf \Pi}(q)}},
\end{eqnarray}
where $q=(i\nu_n,{\bf q})$ with $\nu_n=2\pi nT$ being the boson Matsubara frequency, and ${\bf \Pi}(q)$ is the polarization matrix given by
\begin{eqnarray}
 {\bf \Pi}_{mn}(q)=\int\frac{d^4k}{(2\pi)^4}{\rm Tr}\left[\Gamma_m {\cal S}(k+q) \Gamma_{n}^* {\cal S}(k)\right]
\label{Polarization}
\end{eqnarray}
with the mesonic vertex $\Gamma_m$ defined as
\begin{eqnarray}
  \Gamma_m = \left\{
  \begin{array}{l l}
    \lambda_1^- & \quad m=a^+\\
    \lambda_1^+ & \quad m=a^-\\
    \lambda_3   & \quad m=a_0\\
    \lambda_4^- & \quad m=\kappa^+\\
    \lambda_4^+ & \quad m=\kappa^-\\
    \lambda_6^- & \quad m=\kappa^0\\
    \lambda_6^+ & \quad m=\bar\kappa^0\\
    \lambda_0 & \quad m=\sigma_0\\
    \lambda_8 & \quad m=\sigma_8\\
    i\gamma_5\lambda_1^- & \quad m=\pi^+\\
    i\gamma_5\lambda_1^+ & \quad m=\pi^-\\
    i\gamma_5\lambda_3   & \quad m=\pi_0\\
    i\gamma_5\lambda_4^- & \quad m=K^+\\
    i\gamma_5\lambda_4^+ & \quad m=K^-\\
    i\gamma_5\lambda_6^- & \quad m=K^0\\
    i\gamma_5\lambda_6^+ & \quad m=\bar K^0\\
    i\gamma_5\lambda_0 & \quad m=\eta_0\\
    i\gamma_5\lambda_8 & \quad m=\eta_8\\
  \end{array} \right.
\end{eqnarray}
After the analytical continuation $i\nu_n\rightarrow \omega+i0^+$, the meson masses are determined by the poles of the meson
propagator ${\bf D}(\omega,{\bf q})$ at vanishing momentum ${\bf q}$, i.e.,
\begin{eqnarray}
\det[{\bf 1}-2{\bf G}{\bf \Pi}(\omega=M_m, {\bf q}=0)]=0.
\end{eqnarray}

In the presence of pion condensate $\phi_{ud}$, there arise some off-diagonal elements in the matrices ${\bf G}$ and ${\bf \Pi}$
that do not appear for $\phi_{ud}=0$. We define a matrix ${\bf M}\equiv {\bf 1}-2{\bf G\Pi}$ and denote its elements as
$M(i,j)={\bf M}_{ij}$. After some matrix algebra, we find that the determinant of ${\bf M}$ can be divided into some blocks,
\begin{eqnarray}
\det{\bf M}={\cal C}_1{\cal C}_2{\cal C}_3{\cal C}_4{\cal C}_5{\cal C}_6{\cal C}_7{\cal C}_8,
\label{meson_modes}
\end{eqnarray}
where ${\cal C}_i$ are given by
\begin{eqnarray}
&&{\cal C}_1=\det\left(\begin{array}{cccc}
M(8,8)&  M(8,9)&  M(8,10)&  M(8,11)\cr
M(9,8)&  M(9,9)&  M(9,10)&  M(9,11)\cr
M(10,8)& M(10,9)& M(10,10)& M(10,11)\cr
M(11,8)& M(11,9)& M(11,10)& M(11,11)\end{array}\right),\nonumber\\
&&{\cal C}_2=\det\left(\begin{array}{cccc}
M(17,17)&  M(17,18)&  M(17,1)&  M(17,2)\cr
M(18,17)&  M(18,18)&  M(18,1)&  M(18,2)\cr
M(1,17)&  M(1,18)&  M(1,1)&  M(1,2)\cr
M(2,17)&  M(2,18)&  M(2,1)&  M(2,2)\end{array}\right),\nonumber\\
&&{\cal C}_3=\det\left(\begin{array}{cc}
M(13,13)&  M(13,6)\cr M(6,13)&   M(6,6)\end{array}\right),\nonumber\\
&&{\cal C}_4=\det\left(\begin{array}{cc}
M(15,15)&  M(15,4)\cr M(4,15)&   M(4,4)\end{array}\right),\nonumber\\
&&{\cal C}_5=\det\left(\begin{array}{cc}
M(16,16)&  M(16,5)\cr M(5,16)&   M(5,5)\end{array}\right),\nonumber\\
&&{\cal C}_6=\det\left(\begin{array}{cc}
M(14,14)&  M(14,7)\cr M(7,14)&   M(7,7)\end{array}\right),\nonumber\\
&&{\cal C}_7=M(3,3),\nonumber\\
&&{\cal C}_8=M(12,12).\label{mixing}
\end{eqnarray}
Other matrix elements of ${\bf M}$ that do not appear above vanish automatically.

In the absence of pion condensate $\phi_{ud}$, the nonvanishing off-diagonal elements are $M(8,9)=M(9,8)$ and $M(17,18)=M(18,17)$,
which correspond to the $\sigma_0-\sigma_8$ mixing and $\eta_0-\eta_8$ mixing, respectively. At zero temperature and density, we
recover the meson spectroscopy of three-flavor QCD. The $\eta_0-\eta_8$ mixing and $\sigma_0-\sigma_8$ mixing lead to the real
mesonic excitations, the pseudoscalar $\eta,\eta^\prime$ mesons and scalar $\sigma,f_0$ mesons, respectively.

In the presence of pion condensate $\phi_{ud}\neq0$, all the off-diagonal elements shown in Eq. (\ref{mixing}) are nonzero.
Therefore, the pion condensation generally leads to the mixing between the scalar and pseudoscalar mesons. First, from the expressions of
${\cal C}_7$ and ${\cal C}_8$, we observe that the $\pi_0$ and $a_0$ are the eigenmodes of the mesonic excitations even in the presence
of pion condensation. For other mesonic excitations, we have scalar-pseudoscalar mixing. In detail, the $4\times4$ matrix ${\cal C}_1$
corresponds to the mixing between the scalar $\sigma_0-\sigma_8$ sector and the pseudoscalar $\pi^+-\pi^-$ sector.
${\cal C}_2$ corresponds to the mixing between the scalar $a^+-a^-$ sector and the pseudoscalar $\eta_0-\eta_8$ sector. The $2\times2$ matrices
${\cal C}_3$, ${\cal C}_4$, ${\cal C}_5$, and ${\cal C}_6$ describe the $K^+-\kappa^0$, $K^0-\kappa^+$, $\bar{K^0}-\kappa^-$, and $K^--\bar\kappa^0$ mixing, respectively. Since we consider the case of zero strangeness chemical potential $\mu_{\rm S}=0$, we have
\begin{eqnarray}
{\cal C}_3={\cal C}_5,\ \ \ {\cal C}_4={\cal C}_6.
\end{eqnarray}
Therefore, the mesonic excitations from $K^+-\kappa^0$ and $\bar{K^0}-\kappa^-$ mixing are degenerate, and
the excitations from $K^0-\kappa^+$ and $K^--\bar{\kappa}^0$ mixing are also degenerate.

Now we turn to our numerical results. We focus on the following three cases: (A) $\mu_{\rm I}=0$, $T\neq 0$, (B) $\mu_{\rm I}\neq0$, $T=0$, and
(C) $\mu_{\rm I}\neq 0$, $T\neq 0$. Here we emphasize two points before we present the numerical results: (1) In the superfluid phase with $\phi_{ud}\neq0$, only the $\pi_0$ and $a_0$ mesons carry the same quantum numbers as those in the vacuum; other mesonic excitations are generally mixtures of scalar and pseudoscalar contents. Therefore, we may name the mesonic excitations in the superfluid phase by using the mesons in the normal phase according to the fact that their masses are generally continuous at the superfluid phase transition. (2) Some mesonic excitations have nonzero decay width at high temperature and chemical potential. Here we focus on their real part masses.

\subsection{$\mu_{\rm I}=0$, $T\neq 0$}
At finite temperature and vanishing isospin chemical potential, pion condensation does not occur.
Therefore, we have only the $\eta_0-\eta_8$ mixing and $\sigma_0-\sigma_8$ mixing induced by the
${\rm U}_{\rm A}(1)$ anomaly. In this paper, we consider the case that the ${\rm U}_{\rm A}(1)$
strength $K$ does not depend on the temperature. While the mixing persists to exist at finite
temperature, the meson masses and the mixing angles change with increasing temperature.

\begin{figure}[hbt]
\centering
\includegraphics[width=8.3cm]{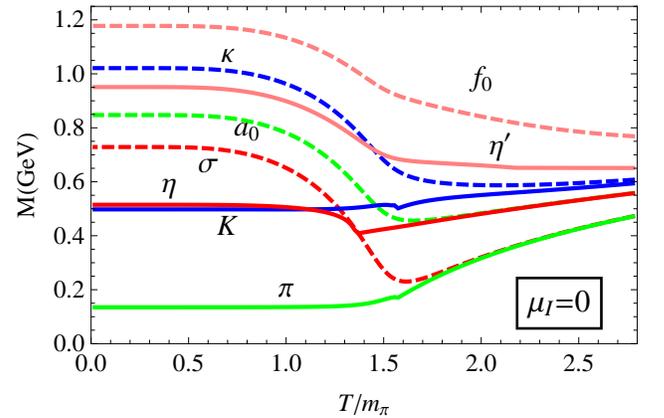}
\caption{(color-online). The meson masses as functions of temperature $T$ at zero
isospin chemical potential.}
\label{mesonmassT}
\end{figure}

In Fig. \ref{mesonmassT}, we show the meson masses with increasing temperature $T$ at $\mu_{\rm I}=0$.
Due to vanishing chemical potentials, the masses of $\pi^+$, $\pi^-$ and $\pi_0$ are degenerate
(denoted by $\pi$), the masses of $K^+$, $K^-$, $K^0$ and $\bar{K}^0$ are degenerate (denoted by $K$),
and the masses of $\kappa^+$, $\kappa^-$, $\kappa^0$ and $\bar{\kappa}^0$ are degenerate (denoted by
$\kappa$). Our results are consistent with those reported in the previous studies~\cite{Renberg,Costa}.
From the results of the chiral condensates shown in Fig. \ref{condensate01}, we observe that the
SU$_{\rm A}(2)$ chiral symmetry for light quarks becomes approximately restored at high temperature.
This fact is also reflected in the meson mass spectra. At high enough temperature, we find that the
masses of the SU$(2)$ chiral partners ($\pi\text{-}\sigma$, $K\text{-}\kappa$,
and $a\text{-}\eta$) become approximately degenerate.

\begin{figure}[hbt]
\centering
\includegraphics[width=7.5cm]{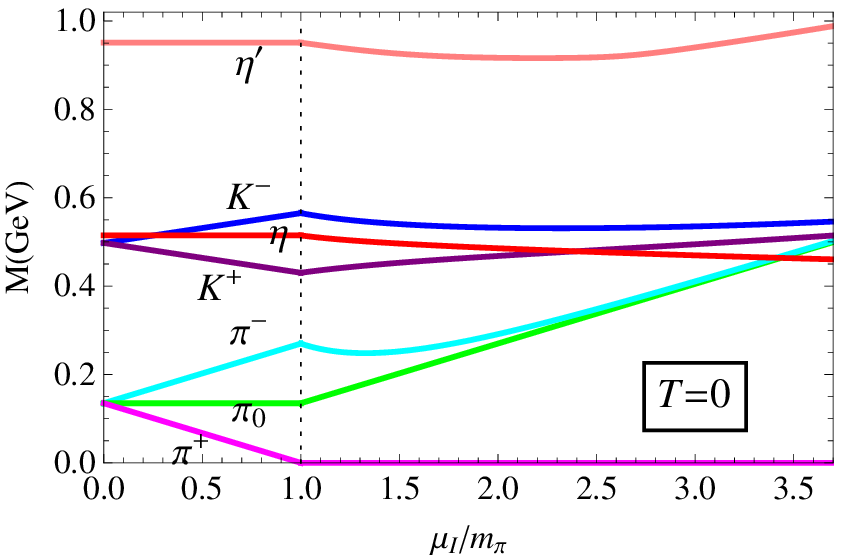}
\includegraphics[width=7.5cm]{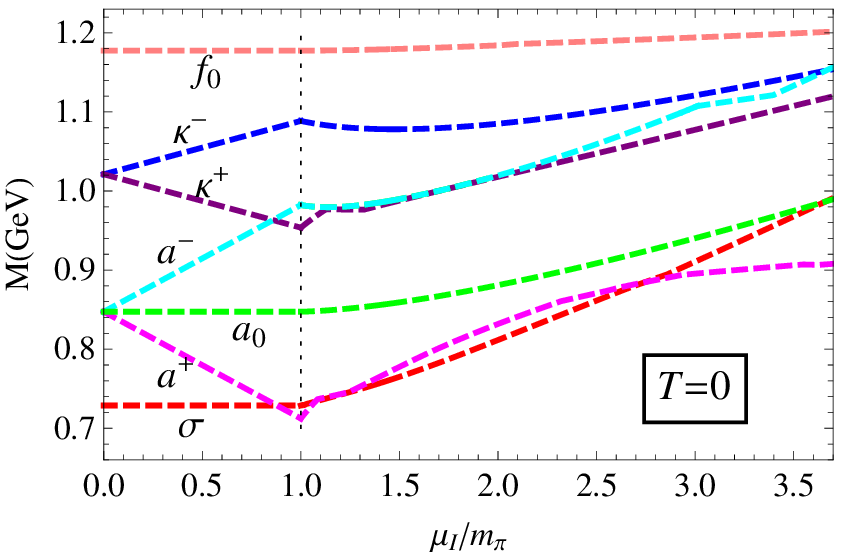}
\caption{(color-online). The meson mass spectra at finite isospin chemical potential
and zero temperature.} \label{mesonmassmuI}
\end{figure}

\subsection{$\mu_{\rm I}\neq0$, $T=0$}

At zero temperature, the system undergoes a quantum phase transition from the vacuum
to the pion condensed phase at $\mu_{\rm I}=m_\pi$. It is interesting to study the behavior
mesonic excitations across this phase transition. Since we consider vanishing strangeness
chemical potential $\mu_{\rm S}$, the mesonic excitations from $K^+-\kappa^0$ and
$\bar{K^0}-\kappa^-$ mixing are degenerate and the excitations from $K^0-\kappa^+$ and
$K^--\bar{\kappa}^0$ mixing are also degenerate. Therefore, in the following we only show
the masses of $K^\pm$ and $\kappa^\pm$ mesons. The numerical results for the meson masses
as functions of the isospin chemical potential are shown in Fig. \ref{mesonmassmuI}.

In the regime $\mu_{\rm I}<m_\pi$, pion condensation does not occur and we have $\phi_{ud}=0$.
At zero temperature, the mesonic excitations are the same as those in the vacuum, except for
the fact that the pole masses are shifted by the isospin chemical potential. For a meson that
carries an isospin quantum number $I$, its mass at finite isospin chemical potential reads
\begin{eqnarray}
M_{I}=M_{\rm v}-I\mu_{\rm I},\ \ \ \ \mu_{\rm I}<m_\pi,
\end{eqnarray}
where $M_{\rm v}$ is its mass in the vacuum. This result can be understood by the fact that
its energy-momentum dispersion at finite isospin chemical potential reads $\omega({\bf q})=
\sqrt{{\bf q}^2+M_{\rm v}^2}-I\mu_{\rm I}$ and its mass is defined as the pole $M_I=\omega({\bf 0})$
at zero momentum. We note that since the pions are the lightest meson in the meson spectra, the $\pi^+$
mass first drops down to zero at $\mu_{\rm I}=m_\pi$. This is consistent with the fact that the charged
pions rather than other mesons start to condense at $\mu_{\rm I}=m_\pi$.

In the superfluid phase $\mu_{\rm I}>m_\pi$, there exists not only the mixing induced by the U$_{\rm A}(1)$
anomaly but also the scalar-pseudoscalar mixing induced by the nonvanishing pion condensate $\phi_{ud}\neq0$.
The masses of the mesonic excitations in the superfluid phase should be determined by solving the equations,
\begin{eqnarray}
{\cal C}_i(\omega=M_m,{\bf q}=0)=0, \ \ \ \ i=1,2,...,8.
\end{eqnarray}
First, at finite isospin chemical potential $\mu_{\rm I}\neq0$, the flavor symmetry for the light quarks,
SU$_{\rm L}(2)\times$SU$_{\rm R}(2)$, is explicitly broken down to U$_{\rm L}(1)\times$U$_{\rm R}(1)\equiv$
U$_{\rm I}(1)\times$U$_{\rm IA}(1)$ with the generator being the third component of the isospin ($I_3$). We note
that the residual U$_{\rm I}(1)$ symmetry is exact even for nonvanishing current quark mass $m_q$. The
U$_{\rm I}(1)$ symmetry is spontaneously broken by the nonzero pion condensate $\phi_{ud}\neq0$, leading to
Bose-Einstein condensation of pions. Therefore, we expect that a massless mesonic excitation appears in the
meson spectra, corresponding to the Goldstone mode in a superfluid. Using the gap equations for the condensates
$\sigma_q$, $\sigma_s$, and $\phi_{ud}$, we find analytically that
\begin{eqnarray}
{\cal C}_1(\omega=0,{\bf q}=0)=0.
\end{eqnarray}
Therefore, our treatment of the condensates and the collective excitations self-consistently results in a
massless mode, i.e., the Goldstone mode. The masses of the other mesonic excitations can only be solved
numerically, except for the $\pi_0$ mass. For the $\pi_0$ meson, we have analytically $M_{\pi_0}=\mu_{\rm I}$,
which is the same as that predicted by the two-flavor models~\cite{ISOother02,note01}.

The numerical results for the mass spectra are shown in Fig. \ref{mesonmassmuI}. We only denote the names
of the mesonic excitations in the normal phase $\mu_{\rm I}<m_\pi$, where the mesons carry the same quantum
numbers as in the vacuum. In the superfluid phase $\mu_{\rm I}>m_\pi$, only the $\pi_0$ and $a_0$ modes carry
their vacuum quantum numbers. For other mesonic excitations, they generally contain both scalar and pseudoscalar
contents. However, we still name the mesonic excitations in the superfluid phase according to the fact that their
masses are continued with the mesons in the normal phase at the phase transition $\mu_{\rm I}=m_\pi$. At finite
isospin chemical potential, the SU$_{\rm A}(2)$ chiral symmetry for the light quarks is explicitly
broken down to U$_{\rm IA}(1)$. Since the chiral condensates for light quarks become very small at large isospin
chemical potential, we may expect approximate restoration of the chiral symmetry in the superfluid phase. Actually,
at large isospin chemical potential (e.g., $\mu_{\rm I}=3m_\pi$), we find that the content of the ``$\pi^-$ meson"
is almost that of the $\sigma$ meson. Therefore, the degeneracy of the masses of the $\pi_0$ and $\pi^-$ meson
reflects the approximate restoration of the chiral symmetry.

\begin{figure}[hbt]
\centering
\includegraphics[width=7.5cm]{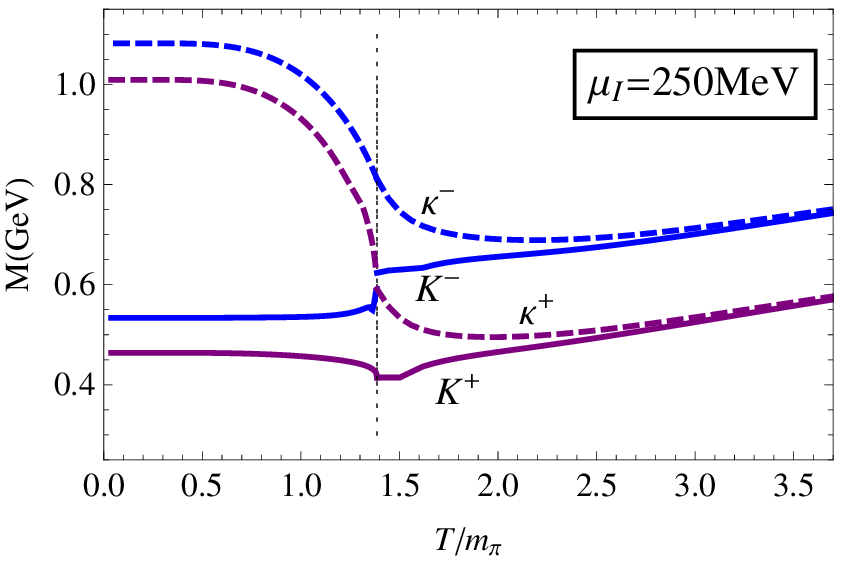}
\includegraphics[width=7.5cm]{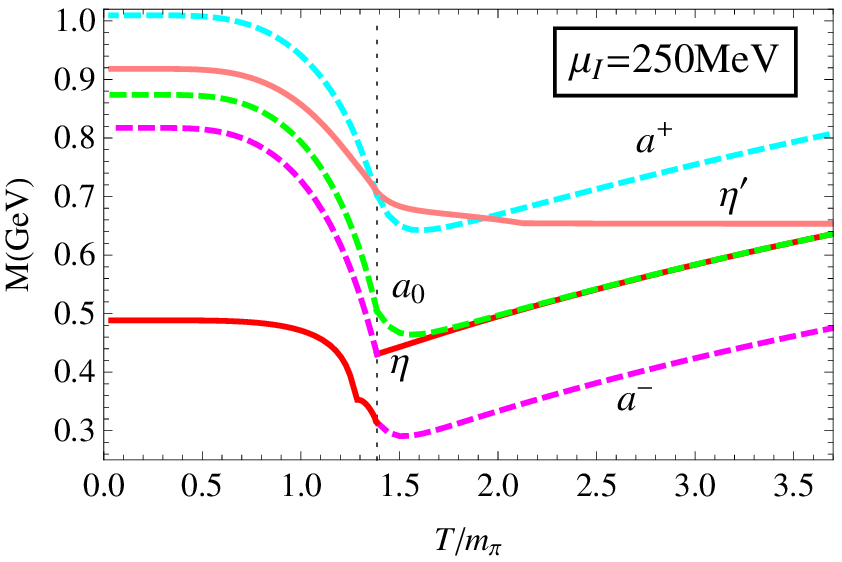}
\includegraphics[width=7.5cm]{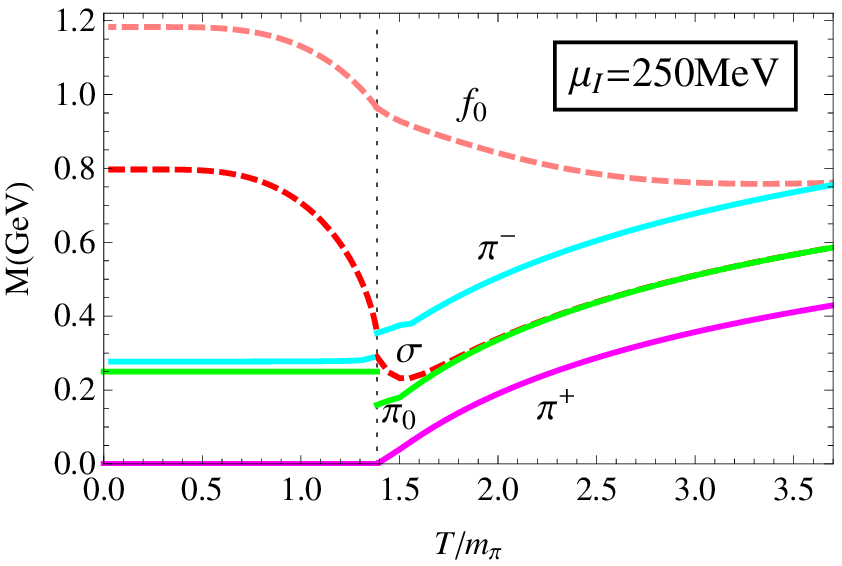}
\caption{(color-online). The meson masses as functions of the temperature $T$ at fixed
isospin chemical potential $\mu_{\rm I}=250$MeV. The thin dashed lines denote the
superfluid phase transition at $T=T_c=187.2$MeV.} \label{mesonmassTmuI}
\end{figure}

\begin{figure}[hbt]
\centering
\includegraphics[width=7.5cm]{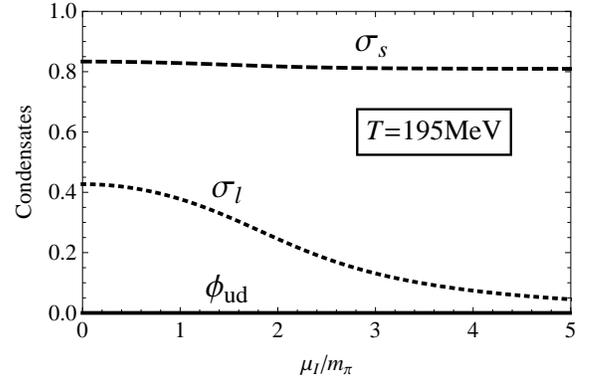}
\caption{The chiral and pion condensates as functions of the isospin chemical potential
$\mu_{\rm I}$ at fixed temperature $T=195$MeV. } \label{condensate05}
\end{figure}

\begin{figure}[hbt]
\centering
\includegraphics[width=7.5cm]{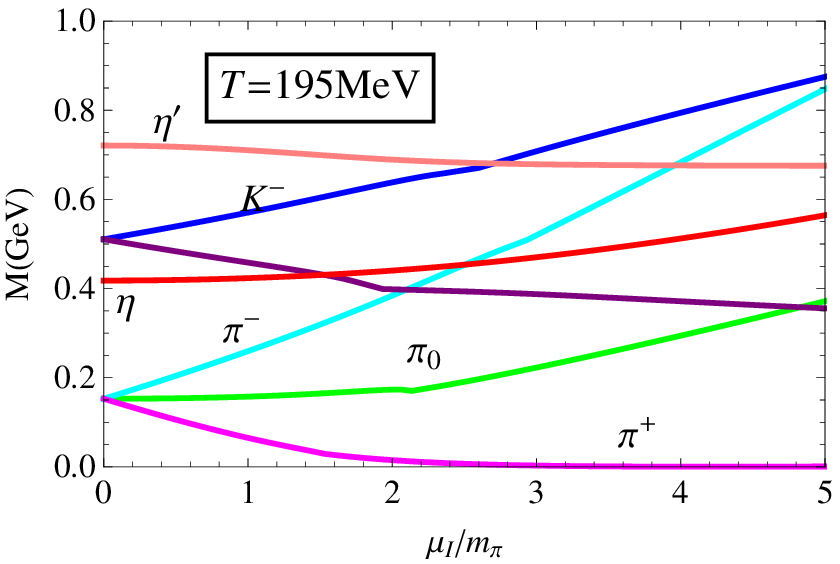}
\includegraphics[width=7.5cm]{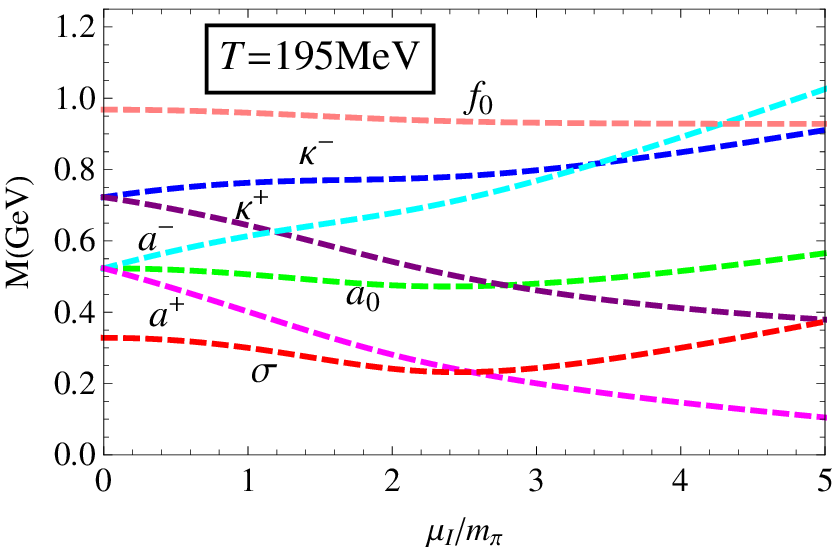}
\caption{(color-online). The meson masses as functions of the isospin chemical potential
$\mu_{\rm I}$ at fixed temperature $T=195$MeV.} \label{mesonmassmuIT}
\end{figure}

\subsection{$\mu_{\rm I}\neq0$, $T\neq 0$}
Finally, we discuss the general case of nonzero temperature and isospin chemical potential.
First, let us fix the isospin chemical potential $\mu_{\rm I}=250$MeV and study the
mesonic excitations across the temperature-induced superfluid phase transition. The
finite-temperature behavior of the chiral and pion condensates is shown in Fig. \ref{condensate04},
which indicates a superfluid phase transition at $T=T_c=187.2$MeV. The finite-temperature
behavior of the meson masses for fixed $\mu_{\rm I}=250$MeV is shown in Fig. \ref{mesonmassTmuI}.

In the superfluid phase $T<T_c$, the mesonic excitations except for $\pi_0$ and $a_0$ contain
both scalar and pseudoscalar contents. We name the excitations according to the fact that their
masses are continuous in the regime $T<T_c$, and we have named the mesonic excitations at $T=0$
in Fig. \ref{mesonmassmuI}. Above the critical temperature $T_c$, the pion condensate $\phi_{ud}$
vanishes and all mesonic excitations can be characterized by their quantum numbers. In
Fig. \ref{mesonmassTmuI}, we observe that some meson masses are discontinuous across the superfluid
phase transition. At high temperature, since the light quark chiral condensate becomes very small,
we expect that the residual U$_{\rm IA}$ chiral symmetry gets restored. Actually, we find that the
masses of the U$(1)$ chiral partners ($K^+\text{-}\kappa^+$, $K^-\text{-}\kappa^-$, $\pi_0\text{-}\sigma$,
and $a_0\text{-}\eta$) become approximately degenerate.

Next we fix the temperature $T=195$MeV and study the isospin splitting effect on the mesonic excitations.
The behavior of the chiral and pion condensates at $T=195$MeV is shown in Fig. \ref{condensate05}. The pion
condensate $\phi_{ud}$ always vanishes and the light quark chiral condensate $\sigma_q$ becomes small at large
isospin chemical potential. Since the pion condensate vanishes, all mesonic excitations can be characterized by their
quantum numbers.  The meson mass spectra are shown in Fig. \ref{mesonmassmuIT}. We find clearly that the
mass splitting of the isospin partners ($K^+-K^-$, $\kappa^+-\kappa^-$, $\pi^+-\pi^-$ and $a^+-a^-$) becomes
larger and larger with increasing isospin chemical potential.

\section{Topological susceptibility at finite isospin density}\label{sec4}
The topological susceptibility $\chi$ is a fundamental correlation function in QCD and is the key to
understanding much of the distinctive dynamics in the U$_{\rm A}(1)$ channel. In this section we
calculate the topological susceptibility $\chi$ at finite isospin chemical potential $\mu_{\rm I}$
within the framework of the three-flavor NJL model. The general expression of $\chi$ to the leading order
in the $1/N_c$ expansion at finite temperature and vanishing chemical potentials for the NJL model
has been derived by Fukushima \emph{et al.}~\cite{Fukushima}. On the other hand, the finite-temperature
behavior of the topological susceptibility can be determined by lattice simulations of QCD at $T\neq0$~\cite{ChiL}.
Therefore, the temperature dependence of the U$_{\rm A}(1)$ anomaly strength $K$ in the NJL model can be
determined by using the lattice data. A vanishingly small $K$ at high temperature can be regarded as
the effective restoration of the U$_{\rm A}(1)$ symmetry.

Since there are no lattice data for the topological susceptibility at finite $\mu_{\rm I}$, in this section
we study the behavior of $\chi$ at finite $\mu_{\rm I}$, by treating the U$_{\rm A}(1)$ anomaly strength $K$
as a constant. The $\mu_{\rm I}$ dependence of the U$_{\rm A}(1)$ anomaly strength $K$ may be determined if
lattice data for $\chi$ at finite $\mu_{\rm I}$ are available in the future. In the following we first derive
a general expression of $\chi$ to the leading order in the $1/N_c$ expansion at finite isospin chemical
potential and temperature, where the pion condensate $\phi_{ud}$ can be nonzero. To this end, we first show
the definition of $\chi$ and therefore start with the QCD Lagrangian density,
\begin{eqnarray}
 {\cal L}_{\text{QCD}}=-\frac{1}{4}F_{\mu\nu}^aF^{a\mu\nu}
+\bar\psi(i\gamma_\mu D^\mu-\hat{m}_0)\psi+\theta Q
\end{eqnarray}
where $F_{\mu\nu}^a$ represents the gluon field strength tensor, $D_\mu=\partial_\mu+igA_\mu$ is the covariant
derivative with $A_\mu$ being the gluon field and $g$ the QCD coupling constant, $\theta$ is the QCD vacuum angle,
and $Q$ is the topological charge density. The topological charge density $Q$ is given by
\begin{eqnarray}
 Q(x)=\frac{g^2}{32\pi^2}F_{\mu\nu}^a\widetilde{F}^{a\mu\nu}.
\end{eqnarray}
The topological susceptibility $\chi$ can be defined as a second-order derivative of the vacuum
energy density $\varepsilon$ with respect to $\theta$ at $\theta=0$,
\begin{eqnarray}
 \chi=\frac{\partial^2\varepsilon}{\partial\theta^2}\Bigg|_{\theta=0}
=\int d^4x\langle 0|{\rm T}Q(x)Q(0)|0\rangle_{\rm connected},
\end{eqnarray}
where T denotes the time-ordering operator, and the subscript ``connected" means to pick out the
diagrammatically connected contributions. On the other hand, $\chi$ can also be regarded as
the zero frequency mode of the Fourier transform of the correlation function
$\langle{\rm T}Q(x)Q(0)\rangle$,
\begin{eqnarray}
\chi=\lim_{k\rightarrow0}\int d^4x e^{-ik\cdot x}\langle{\rm T}Q(x)Q(0)\rangle.
\end{eqnarray}

In order to calculate $\chi$ in the NJL model, we need to find a correspondent to $Q(x)$ in the model.
In QCD the axial current $J_5^\mu=\bar\psi\gamma^5\gamma^\mu\psi$ is not conserved due to the
U$_{\rm A}(1)$ anomaly induced by the instanton effect. This fact can be expressed as
\begin{eqnarray}
\partial_\mu J_5^\mu(x)=2N_fQ(x)+2i\bar\psi m\gamma_5\psi.
\label{Q_QCD}
\end{eqnarray}
On the other hand, in the NJL model, the U$_{\rm A}(1)$ anomaly is caused by the KMT term. We find~\cite{Fukushima}
\begin{eqnarray}
\partial_\mu J_5^\mu=4N_fK\text{Imdet}\Phi+2i\bar\psi m\gamma_5\psi,
\label{Q_NJL}
\end{eqnarray}
where $\Phi_{ij}=\bar{\psi}_i(1-\gamma_5)\psi_j$ with $i,j$ being the flavor indices. Comparing the
expressions (\ref{Q_QCD}) and (\ref{Q_NJL}), we find that the topological charge density in the NJL model can be
defined as
\begin{eqnarray}
Q(x)=2K\text{Imdet}\Phi=-iK\left[\text{det}\Phi-(\text{det}\Phi)^*\right].
\end{eqnarray}

Having obtained the expression of the topological charge density $Q(x)$ in the NJL model, we can go on to calculate
the topological susceptibility $\chi$ according to the definition (52). Using the identities
\begin{eqnarray}
 \textup{det}\Phi=\epsilon^{ijk}
(\bar\psi_u\Gamma_-\psi_i)(\bar\psi_d\Gamma_-\psi_j)(\bar\psi_s\Gamma_-\psi_k),\nonumber\\
 (\textup{det}\Phi)^*=\epsilon^{lmn}
(\bar\psi_u\Gamma_+\psi_l)(\bar\psi_d\Gamma_+\psi_m)(\bar\psi_s\Gamma_+\psi_n),
\end{eqnarray}
where $\Gamma_\pm=1\pm\gamma_5$ and the summation over the flavor indices ($i,j,k,l,m,n$) is implicit,
we can express the topological susceptibility $\chi$ as~\cite{Fukushima}
\begin{widetext}
\begin{eqnarray}
\chi&=&\int d^4 x\langle 0|\text{T}Q(x)Q(0)|0\rangle_{\text{connected}}\nonumber\\
&=&-\ K^2\int d^4x\epsilon^{ijk}\epsilon^{lmn}
\langle 0|\text{T}\lbrace
(\bar\psi_u\Gamma_-\psi_i)(\bar\psi_d\Gamma_-\psi_j)(\bar\psi_s\Gamma_-\psi_k)(x)
(\bar\psi_u\Gamma_-\psi_l)(\bar\psi_d\Gamma_-\psi_m)(\bar\psi_s\Gamma_-\psi_n)(0) \nonumber\\
&&-\ (\bar\psi_u\Gamma_-\psi_i)(\bar\psi_d\Gamma_-\psi_j)(\bar\psi_s\Gamma_-\psi_k)(x)
(\bar\psi_u\Gamma_+\psi_l)(\bar\psi_d\Gamma_+\psi_m)(\bar\psi_s\Gamma_+\psi_n)(0) \nonumber\\
&&-\ (\bar\psi_u\Gamma_+\psi_i)(\bar\psi_d\Gamma_+\psi_j)(\bar\psi_s\Gamma_+\psi_k)(x)
(\bar\psi_u\Gamma_-\psi_l)(\bar\psi_d\Gamma_-\psi_m)(\bar\psi_s\Gamma_-\psi_n)(0) \nonumber\\
&&+\ (\bar\psi_u\Gamma_+\psi_i)(\bar\psi_d\Gamma_+\psi_j)(\bar\psi_s\Gamma_+\psi_k)(x)
(\bar\psi_u\Gamma_+\psi_l)(\bar\psi_d\Gamma_+\psi_m)(\bar\psi_s\Gamma_+\psi_n)(0)\rbrace
|0\rangle_{\text{connected}}.
\end{eqnarray}
\end{widetext}
Next we need to evaluate the above four matrix elements. Following Wick's theorem, we take full contraction in terms
of the quark propagator ${\cal S}(x,x^\prime)$ which has been constructed in Eq. (\ref{Qpropagator}) with the condensates
determined by the self-consistent gap Eqs. (\ref{gapeq}).

\begin{figure}[hbt]
\centering
\includegraphics[width=5cm]{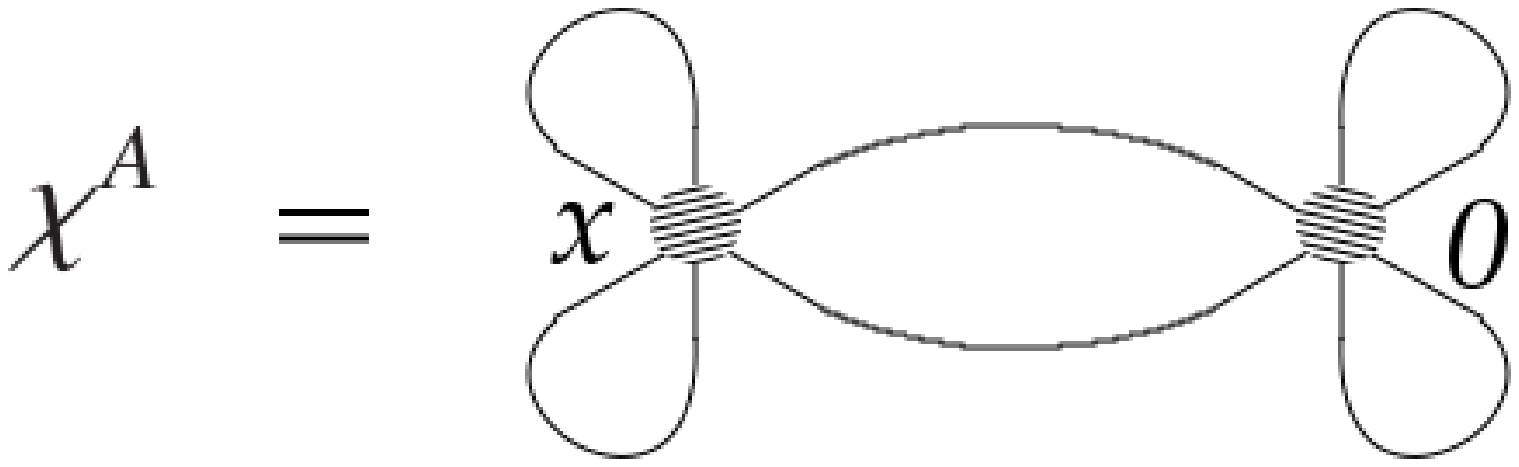}
\caption{The simplest leading-order contribution to the topological susceptibility
in the $1/N_c$ expansion. The gray circle denotes the U$_{\rm A}(1)$ anomaly
coupling $K$. }
\label{feynmann01}
\end{figure}

\begin{figure}[hbt]
\centering
\includegraphics[width=8.5cm]{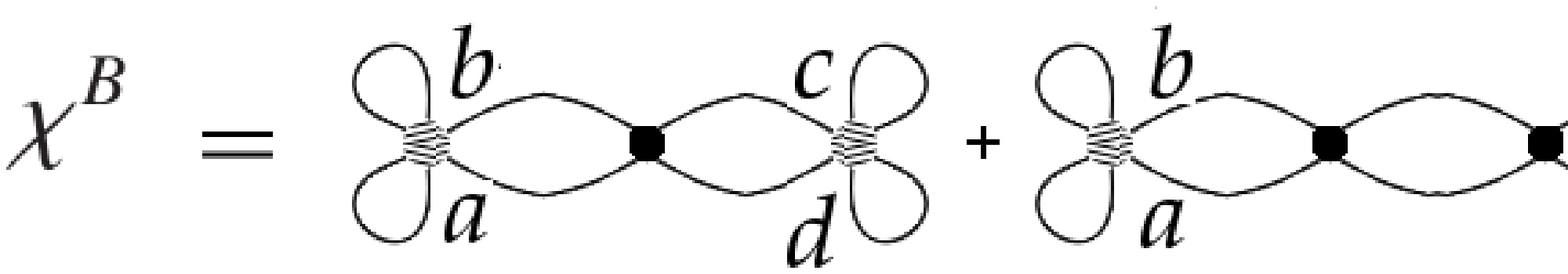}
\includegraphics[width=5cm]{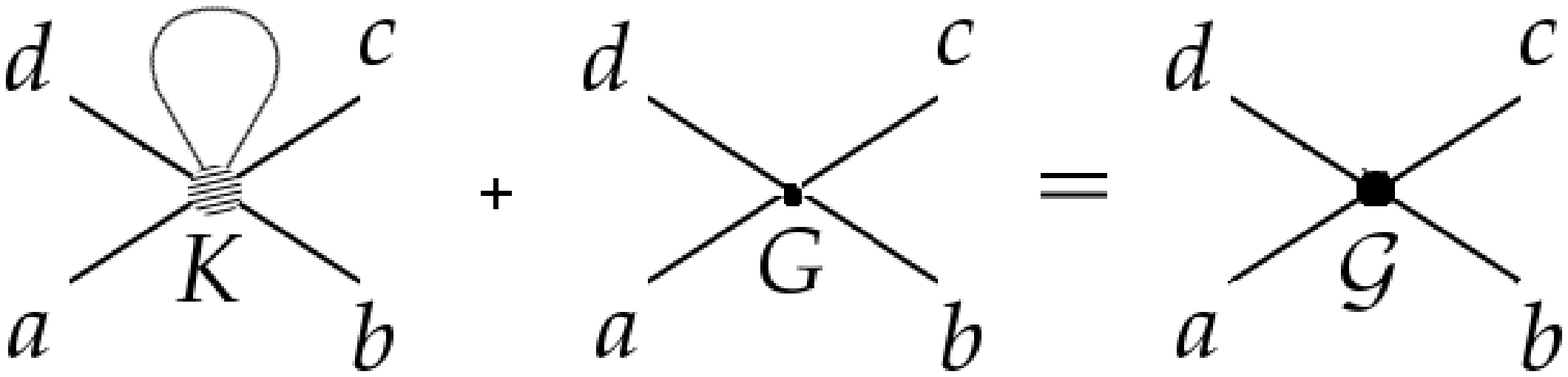}
\caption{Other leading-order contributions (ring diagrams) to the topological
susceptibility. }
\label{feynmann02}
\end{figure}

To be self-consistent, we consider only the leading-order contributions in the $1/N_c$ expansion. The simplest
leading-order contribution $\chi^A\sim O(N_c^5)$ is shown in Fig. \ref{feynmann01}. In the presence of pion condensate
$\phi_{ud}$, the quark propagator ${\cal S}(x,x^\prime)$ is not diagonal in the flavor space. This brings
many more contributions than the case of $\phi_{ud}=0$. To evaluate $\chi^A$, we just need to replace the
topological charge density $Q(x)$ by its mean-field approximation
\begin{eqnarray}
Q_{\rm mf}(x)&\simeq&\left(\frac{1}{2\sqrt{6}}|\phi_{ud}|^2+\frac{4}{\sqrt{6}}\sigma_l\sigma_s
+\frac{2}{\sqrt{6}}\sigma_l^2\right)K(\bar\psi\lambda_0i\gamma_5\psi)\nonumber\\
&+&\left(-\frac{1}{2\sqrt{3}}|\phi_{ud}|^2+\frac{2}{\sqrt{3}}\sigma_l\sigma_s
-\frac{2}{\sqrt{3}}\sigma_l^2\right)K(\bar\psi\lambda_8i\gamma_5\psi)\nonumber\\
&-&\frac{1}{\sqrt{2}}\phi_{ud}^*K\sigma_s(\bar\psi\lambda_1^+\psi)
-\frac{1}{\sqrt{2}}\phi_{ud}K\sigma_s(\bar\psi\lambda_1^-\psi).
\end{eqnarray}
Then the contribution $\chi^A$ can be expressed in a compact form,
\begin{eqnarray}
\chi^A=4K^2P^{\rm T}\Pi P,
\end{eqnarray}
where the matrices $P$ and $\Pi$ are given by
\begin{eqnarray}
P=\left( \begin{array}{ccc}
\frac{-1}{4\sqrt{6}}|\phi_{ud}|^2-\frac{2}{\sqrt{6}}\sigma_l\sigma_s
-\frac{1}{\sqrt{6}}\sigma_l^2  \\
\frac{1}{4\sqrt{3}}|\phi_{ud}|^2-\frac{1}{\sqrt{3}}\sigma_l\sigma_s
+\frac{1}{\sqrt{3}}\sigma_l^2  \\
\frac{1}{2\sqrt{2}}\phi_{ud}^*\sigma_s \\
\frac{1}{2\sqrt{2}}\phi_{ud}\sigma_s \end{array} \right)
\end{eqnarray}
and
\begin{eqnarray}
\Pi=\left( \begin{array}{cccc}
\Pi_{\eta_0\eta_0} & \Pi_{\eta_0\eta_8} & \Pi_{\eta_0a^+} & \Pi_{\eta_0a^-} \\
\Pi_{\eta_8\eta_0} & \Pi_{\eta_8\eta_8} & \Pi_{\eta_8a^+} & \Pi_{\eta_8a^-} \\
\Pi_{a^+\eta_0} & \Pi_{a^+\eta_8} & \Pi_{a^+a^+} & \Pi_{a^+a^-} \\
\Pi_{a^-\eta_0} & \Pi_{a^-\eta_8} & \Pi_{a^-a^+} & \Pi_{a^-a^-} \end{array} \right).
\end{eqnarray}
Note that the matrix $\Pi$ is actually the matrix in the determinant ${\cal C}_2$, but
evaluated at $(\omega,{\bf q})=(0,{\bf 0})$.

\begin{figure}[hbt]
\centering
\includegraphics[width=7.5cm]{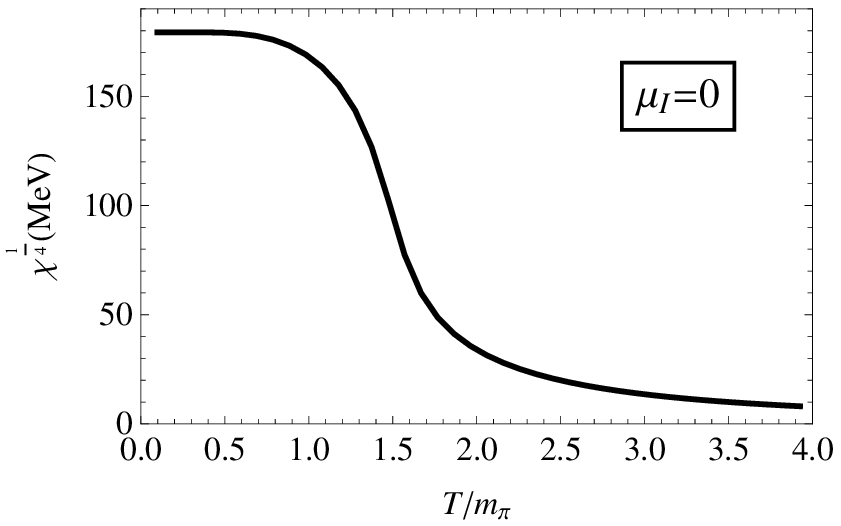}
\includegraphics[width=7.5cm]{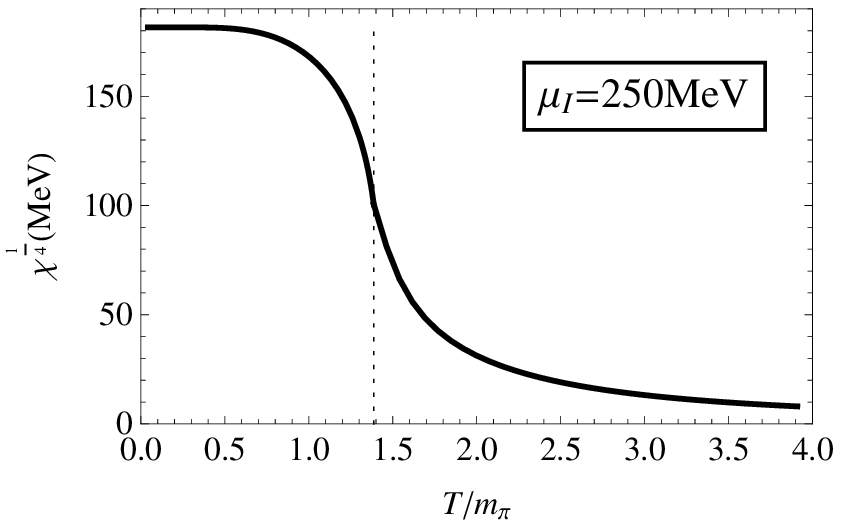}
\caption{The topological susceptibility $\chi$ (we show $\chi^{1/4}$) as a function of
temperature $T$ for $\mu_{\rm I}=0$ and $\mu_{\rm I}=250$MeV. The thin dashed line
denotes the superfluid phase transition.} \label{topoT}
\end{figure}

\begin{figure}[hbt]
\centering
\includegraphics[width=7.5cm]{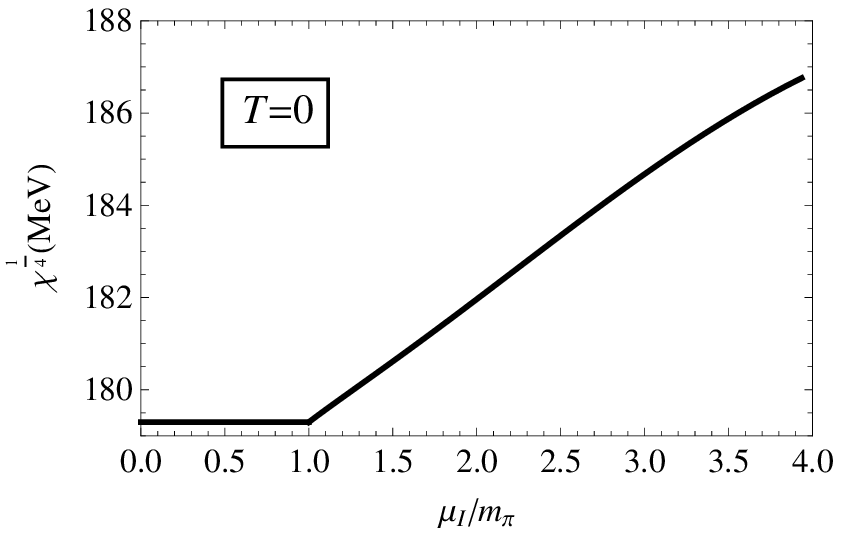}
\includegraphics[width=7.5cm]{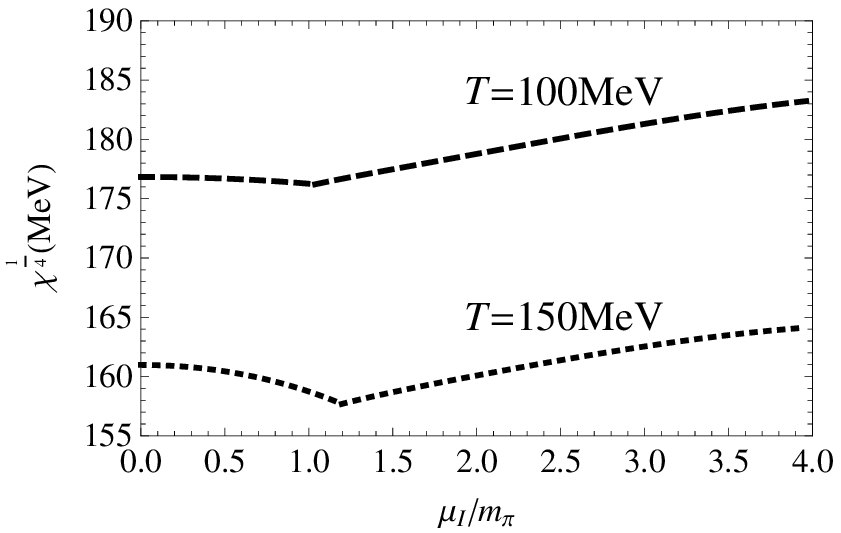}
\caption{The topological susceptibility $\chi$ (we show $\chi^{1/4}$) as a function of
$\mu_{\rm I}$ for $T=0$ and $T\neq0$.} \label{topomuI}
\end{figure}

Other contributions to $\chi$ that are of the same order in $1/Nc$ expansion as $\chi^A$ are shown in Fig. \ref{feynmann02}.
They are regarded as the ring diagrams, and we have to include all these ring diagrams for consistency of the $1/N_c$ expansion.
They are of the same order as Fig. \ref{feynmann01}. To show this, we notice that while each four-point vertex is of order
$O(N_c^{-1})$, it is compensated by a factor $N_c$ coming from its neighboring loop. The sum of these ring diagrams with
the one-loop diagram included can be physically interpreted as the propagations of some certain mesonic modes. In the absence
of the pion condensate, it can be interpreted as the propagations of $\eta_0$ and $\eta_8$ mesons as well as their mixing. However,
for $\phi_{ud}\neq0$, we will see that the propagations of $a^\pm$ mesons as well as their mixing with $\eta_0$ and $\eta_8$ mesons
are involved. Note that the energy and momentum of the propagating mesonic modes are zero; $(\omega,{\bf q})=(0,{\bf 0})$,
which just reflects the fact that $\chi$ is the zero frequency mode of the Fourier transform of $\langle{\rm T}Q(x)Q(0)\rangle$.
Summing all these ring diagrams, we find that the contribution $\chi^B$ can be expressed as
\begin{eqnarray}
\chi^B=4K^2P^{\rm T}\Pi\frac{2{\cal G}}{1-2{\cal G}\Pi}\Pi P,
\end{eqnarray}
where the effective coupling matrix ${\cal G}$ is given by
\begin{eqnarray}
{\cal G}=GI_4+K\left(\begin{array}{cc}-A&\phi_{ud}B^{\rm T} \cr \phi_{ud}B&\frac{1}{2}\sigma_sI_2\end{array}\right).
\end{eqnarray}
Here the matrices $A$ and $B$ have been defined in (39).

The final result for the topological susceptibility reads
\begin{eqnarray}
\chi=\chi^A+\chi^B=4K^2P^{\rm T}\frac{\Pi}{1-2{\cal G}\Pi}P.
\end{eqnarray}
At vanishing isospin chemical potential, $\mu_{\rm I}=0$, by setting the pion condensate $\phi_{ud}=0$, we find that the above result
for $\chi$ recovers the one derived by Fukushima \emph{et al.}~\cite{Fukushima}.

Now we present the numerical results for the topological susceptibility $\chi$ at finite isospin chemical potential and at finite
temperature. For our model parameter set, the topological susceptibility $\chi$ in the vacuum reads $\chi_0=(179.3{\rm MeV})^4$,
consistent with the lattice data~\cite{ChiL}. In Fig. \ref{topoT}, we show the finite-temperature behavior of $\chi$ at vanishing
and at finite isospin chemical potentials. We find that the temperature effect leads to suppression of the topological susceptibility,
both in the normal phase and in the pion superfluid phase. We may expect that the density effect would also reduce the topological
susceptibility. The behavior of $\chi$ with increasing isospin chemical potential $\mu_{\rm I}$ is shown in Fig. \ref{topomuI}.
However, we find that the isospin density effect is quite different from the temperature effect. At zero temperature, $\chi$ keeps
its vacuum value in the regime $\mu_{\rm I}<m_\pi$ and then starts to increase with increasing $\mu_{\rm I}$ in the superfluid
phase $\mu_{\rm I}>m_\pi$, where a nonzero isospin density $n_{\rm I}$ is generated. At finite temperature, $\chi$ first decreases
in the normal phase and reaches a minimum at the superfluid transition. In the superfluid phase, $\chi$ becomes again enhanced by
the isospin chemical potential. However, the increase in the topological susceptibility is not large. At zero temperature, the
quantity $\chi^{1/4}$ is increased by only about $8$MeV at $\mu_{\rm I}=4m_\pi$. The enhancement of the topological susceptibility
at finite isospin density is similar to the behavior of the gluon condensate~\cite{HFiso}, which also gets enhanced by the isospin
density effect.

Finally, we point out that the numerical results presented above are obtained by treating the U$_{\rm A}(1)$ anomaly $K$ as a
constant, which implies no effective restoration of the U$_{\rm A}(1)$ symmetry. However, it is generally believed that the
U$_{\rm A}(1)$ symmetry becomes effectively restored at high density and therefore $K$ is generally $\mu_{\rm I}$ dependent.
The behavior of $\chi$ observed above could be qualitatively reliable at small and even at moderate isospin density, since
we expect the density dependence of $K$ is slight there. At large enough isospin density, we expect $K$ goes to zero
and the U$_{\rm A}(1)$ symmetry gets effectively restored. If the lattice data for $\chi$ at finite $\mu_{\rm I}$ becomes
available in the future, we can determine the $\mu_{\rm I}$ dependence of $K$ and therefore study the restoration of the
U$_{\rm A}(1)$ symmetry at finite density.

\section{summary }\label{sec5}
In summary, we have studied the three-flavor Nambu--Jona-Lasinio model with U$_{\rm A}(1)$
anomaly at finite isospin chemical potential. Similar to the two-flavor models, the three-flavor
NJL model also predicts a quantum phase transition from the vacuum to the pion superfluid phase,
which takes place exactly at $\mu_{\rm I}=m_\pi$. However, due to the U$_{\rm A}(1)$ anomaly,
the strangeness degree of freedom couples to the light quark degrees of freedom,  and the strange
quark effective mass depends on the pion condensate. This coupling is absent if the U$_{\rm A}(1)$
anomaly strength $K$ is turned off. Numerically, we find that the strange quark condensate and
the strange quark effective mass change very slightly in the pion superfluid phase, which verifies
the validity of the two-flavor models. To study the mesonic excitations in the isospin dense medium,
we have constructed the effective four-fermion interaction of the KMT term in the presence of pion
condensation. We find that the pion condensation generally induces scalar-pseudoscalar interaction
in the presence of U$_{\rm A}(1)$ anomaly strength $K$. The propagator of the mesonic excitations
is established within the framework of the random-phase approximation. The meson mass spectra are
studied at finite isospin chemical potential and temperature. Finally, we have derived the general
expression for the topological susceptibility $\chi$ at finite isospin chemical potential with
nonvanishing pion condensate. We find that the topological susceptibility $\chi$ gets enhanced in
the pion superfluid phase; that is, the isospin density effect leads to the enhancement of $\chi$,
in contrast to the finite temperature effect which generally reduces $\chi$. Our model predictions
may be tested by future lattice simulations of $2+1$ flavor QCD at finite isospin density.

{\bf Acknowledgements:} T. X. and P. Z. are supported by the NSFC under Grant No. 11079024 and the MOST
under Grant No. 2013CB922000, and L. H. acknowledges the support from the Helmholtz International
Center for FAIR within the framework of the LOEWE program launched by the state of Hesse.

\end{document}